\documentclass{jfm}
\usepackage{graphicx,rotating}
\usepackage{epstopdf, epsfig}

\usepackage{caption,subcaption}
\usepackage{lscape}
\usepackage{afterpage}
\usepackage{amsmath}
\usepackage{color}

\shorttitle{DNS of turbulent jets}
\shortauthor{Constante-Amores}

\title{Direct numerical simulations of transient turbulent  
jets: vortex-interface interactions}
\shortauthor{C. R. Constante-Amores}
\author[Constante-Amores {\it et al.}]
{C. R. Constante-Amores$^1$, 
\ns L. Kahouadji$^1$,\ns A. Batchvarov$^1$,\ns S. Shin$^2$, \ns J. Chergui$^3$, \ns D. Juric$^3$ and O. K. Matar$^1$
\corresp{\email{o.matar@imperial.ac.uk}}}%

\affiliation{$^1$Department of Chemical Engineering, Imperial College London, South Kensington Campus, London SW7 2AZ, United Kingdom \\[\affilskip]
$^2$Department of Mechanical and System Design Engineering, Hongik University, Seoul 04066, Republic of Korea \\[\affilskip]
$^3$ Universit\'e Paris Saclay, Centre National de la Recherche Scientifique (CNRS), Laboratoire Interdisciplinaire des Sciences du Num\'erique (LISN), 91400 Orsay, France \\[\affilskip]}

\begin{document}

\maketitle

\begin{abstract}
The breakup of an interface into a cascade of droplets and their subsequent coalescence is a generic problem of central importance to a large number of industrial settings such as mixing, separations, and combustion. We study the breakup of a liquid jet introduced through a cylindrical nozzle into a stagnant viscous phase via a hybrid interface-tracking/level-set method to account for the surface tension forces in a three-dimensional Cartesian domain. Numerical solutions are obtained for a range of Reynolds (Re) and Weber (We) numbers. We find that the interplay between the azimuthal and streamwise vorticity components leads to different interfacial features and flow regimes in Re-We space. We show that the streamwise vorticity plays a critical role in the development of the three-dimensional instabilities on the jet surface. In the inertia-controlled regime at high Re and We, we expose the details of the spatio-temporal development of the vortical structures affecting the interfacial dynamics. A mushroom-like structure is formed at the leading edge of the jet inducing the generation of a liquid sheet in its interior that undergoes rupture to form droplets. These droplets rotate inside the mushroom structure due to their interaction with the prevailing vortical structures. Additionally, Kelvin-Helmholtz vortices that form near the injection point deform in the streamwise direction to form hairpin vortices, which, in turn, trigger the formation of interfacial lobes in the jet core. The thinning of the lobes induces the creation of holes which expand to form liquid threads that undergo capillary
breakup to form droplets.
\end{abstract}

\section{Introduction\label{intro_jet_exp}}
The breakup of a dispersed fluid in a stagnant phase is a classical problem in multiphase flows, as it encompasses a multitude of interfacial singularities, which exemplify situations wherein the interface undergoes topological transitions, e.g., liquid-threads breakup into drops, and merging of drops/bubbles due to coalescence. These transitions involve the development of singularities where interfacial distances vanish and velocity fields diverge, and the system is controlled by a combination of capillary, inertial, and viscous forces. The complex topological nature of the jet phenomenon has fascinated the scientific community for decades which has led to numerous, comprehensive reviews, see for example  \citet{Lin_arfm_1998,Eggers_rpp_2008,Lasheras_arfm_2000}.

Jet breakup is influenced by a range of multi-scale physics; these include the interaction of turbulence with interfaces (creating cascades of motion featuring a large separation of scales), capillarity, potentially complex rheology, the presence of fields (e.g. gravitational, electromagnetic), as well as heat transfer and phase change. \cite{reitz_book_1986} proposed four jet breakup regimes depending on the appearance of the jet far downstream from the injection point. In the `Rayleigh regime', the onset of breakup is the Rayleigh-Plateau instability, in which the growth of the linear modes on the jet-surface leads to the formation of large droplets with respect to the jet-nozzle. In the `first/second wind-induced' regime the resulting droplets have roughly the same scale or smaller as the jet-nozzle. Finally, in the `atomisation regime', the generated turbulence helps to create spatio-temporal chaos resulting in droplets which are up to four orders of magnitude smaller than the size of the injection nozzle. The physics of droplet generation in turbulent jets remains partially-understood despite the significant scientific attention it has received over the years   \citep{Dombrowski_ptrsl_1954,Hoyt_jfm_1977,Lasheras_arfm_2000,Marmottant_jfm_2004,Villermaux_prl_2004,Eggers_rpp_2008}.  The rest of this introduction aims to provide an up-to-date summary of the experimental and numerical efforts to study the jet breakup phenomenon.

The  ground-breaking experiments of \cite{Hoyt_jfm_1977} and \citet{Taylor1983} showed the complex topological features of the surface waves formed on the jet. They revealed that these waves are responsible for the transition from laminar to turbulent flow, and found underlying similarities  of the occurring instabilities to inviscid linear theory. \cite{Marmottant_jfm_2004} with their pioneering experiments scrutinised  the various stages of the jet dynamics: from the growth of linear modes (through a Kelvin-Helmholtz instability) that characterises the early-time dynamics, to the development of nonlinearities leading to `primary' and subsequent `secondary' breakup events (through long filament pinchoff modulated by a Rayleigh-Taylor instability), and the formation of a cascade of droplet sizes. These authors found that mean droplet-size is proportional to the wavelength selected during the Rayleigh?Taylor instability (also observed by \citet{varga_lasheras_hopfinger_2003}). In the same premise, \cite{Kooij_prx_2018} showed that the droplet size distribution is also a function of the nozzle geometry and the surrounding pressure. More recently, \citet{ibarra_jfm_2020} presented the results of an experimental study of the spatial evolution of turbulent immiscible liquid-liquid jets; however, a detailed account of the spatio-temporal development, and critical mechanisms leading to droplet generation remains outstanding.

The multi-scale nature of the flow, and the complex interfacial topology complicate the experimental scrutiny of the different physical mechanisms occurring across the scales. Thus, elucidating the fundamental physics of this problem has also relied on high-fidelity simulations exemplified by the work of \citet{Bianchi_2005, Bianchi_2007,Menard_ijmf_2007,Desjardins_jcp_2008,Gorokhovski_arfm_2008,Desjardins_as_2010,Shinjo_2010,Hermann_as_2011,Chenadec_phd_2012,Desjardins_as_2013,Jarrahbashi_jfm_2016,Ling_prf_2017,Agbaglah_jfm_2017,Zandian_jfm_2018,Ling_jfm_2019,zandian_sirignano_hussain_2019}. The first step towards the use of numerical simulations to understand the physical mechanisms at play was conducted by \cite{Desjardins_as_2010}, who were able to identify a sequence of essential steps during the spatio/temporal interfacial development of a planar liquid jet-segment: the formation of initial corrugations on the surface, followed by the development of ligaments whose capillary instability leads to droplet formation. They also showed that the early interfacial corrugations are a consequence of the turbulent eddies which carry enough kinetic energy to overcome capillary forces.

Using a similar approach to \cite{Desjardins_as_2010}, but for a cylindrical liquid jet-segment surrounded by an outer gas phase, \cite{Jarrahbashi_jfm_2016} provided a comprehensive study of the flow structures in terms of the vortex-surface interaction. They showed that `hole formation' of a liquid-sheet is an essential requirement to trigger the formation of droplets, and the thinning of the liquid-sheet is driven by the superposition of hairpin-vortices near the interface rather than by capillarity action. Similar findings have been reported for a planar liquid jet-segment surrounded by an outer gas phase \cite{Zandian_jfm_2018}, and for the transient dynamics of a cylindrical liquid jet surrounded by a coaxial air phase \citet{zandian_sirignano_hussain_2019}. \cite{Ling_prf_2017,Ling_jfm_2019} performed  simulations of a two-phase mixing layer between parallel gas  and liquid streams to investigate the interfacial dynamics, and the statistics for the multiphase turbulence. They also observed that the formation of ligaments, and subsequently-formed droplets, are triggered by the hole-induced perforation of the liquid sheets. 

In the previous numerical studies, interface-capturing capabilities were used to account for the surface tension forces in the absence of intermolecular forces (i.e., disjoining pressure). For static sheets, the disjoining pressure  is neglected as the minimum computational cell is larger than the film sheet thickness in which the intermolecular forces will drive its  perforation. Therefore, the hole formation is an outcome of the numerical cut-off interfacial length scale, i.e., minimum mesh size ($O(10^{-6})$m). Nevertheless, recent experiments \citep{Kooij_prx_2018,marston_jfm_2016,neel_villermaux_2018} demonstrate the existence of hole  formation in dynamic sheets with a characteristic film thickness on the order of microns. 

In this study, we aim to provide a comprehensive explanation of the physical mechanisms governing the interfacial dynamics of turbulent jets focusing on the less well-studied liquid-liquid systems. We will perform high-resolution three-dimensional direct numerical simulations using a hybrid interface-tracking/level-set approach to resolve the interfacial dynamics. We will demonstrate how the interaction between the vortical structures, which accompany the development of the flow, and the interface influence the mechanisms underlying droplet generation over a wide range of Reynolds and Weber numbers. 

The rest of this paper is organised as follows. In Section \ref{sec:numerical_methods}, we present the governing equations along with the numerical technique used to carry out the simulations. In Section \ref{sec:Results}, we begin with the presentation of a regime map in $Re-We$ space, classifying jet spatio-temporal development, followed by an in-depth discussion of the vortex-surface interactions linked to the topological changes; in addition, we elucidate the role of hole formation as a precursor to droplet generation. Finally, concluding remarks are provided in Section \ref{conclu}.

\begin{figure}
 \includegraphics[width=1\linewidth]{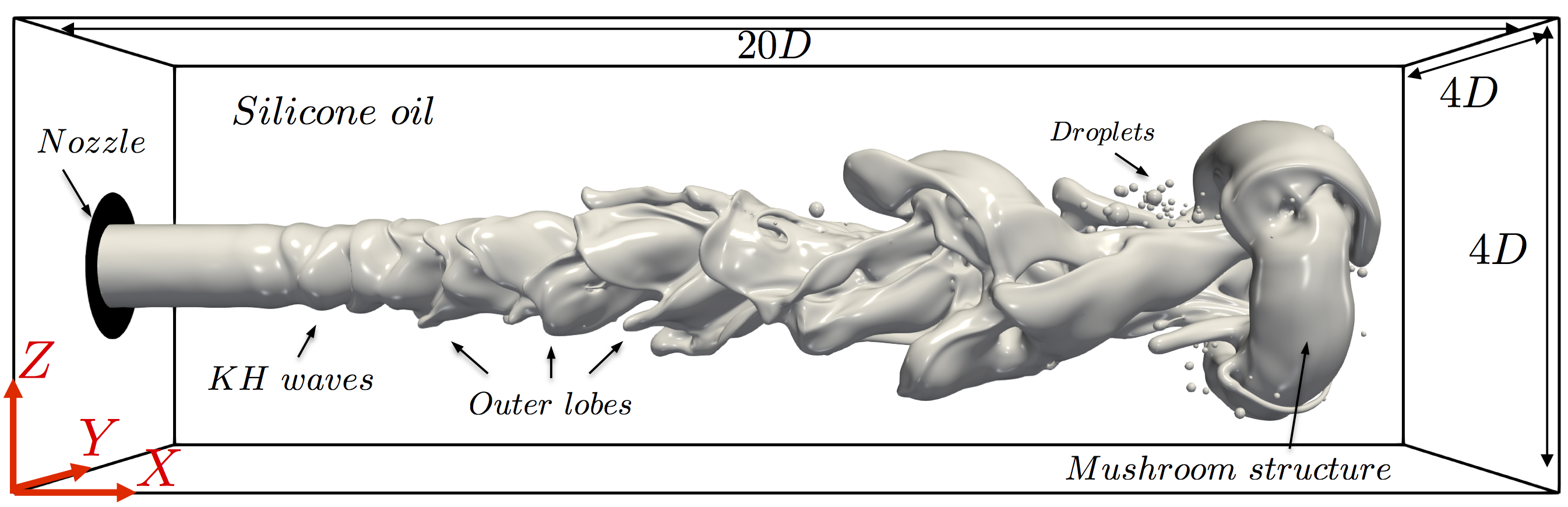}
 \caption{ \label{fig:introductory_image} Injection of a turbulent water jet into a stagnant silicone-oil phase. Three-dimensional representation of the interfacial shape for Case 4 of table \ref{tab:cases} (i.e., $Re=6530$ and $We=303$) at $t=28.97$ with definitions of particular regions and specific features discussed in this work.}
\end{figure}

\section{Problem formulation and numerical techniques\label{sec:numerical_methods}}

Figure \ref{fig:introductory_image} shows a three-dimensional representation of the interface highlighting several specific features that are discussed in detail in the present work: Kelvin Helmholtz (KH) waves close to the injection nozzle, outer lobes formed on the main body of the jet, a leading edge `mushroom'-like structure, and droplets resulting from the atomisation process. The numerical framework is based on solving the two-phase incompressible Navier-Stokes equations in a three-dimensional Cartesian  domain $x = (x, y, z)$.  The surface tension force in the momentum equation is treated by using a hybrid front-tracking/level-set method presented by \citet{Shin_jcp_2002}. The governing flow equations in the `one-fluid' formulation are described by

\begin{equation}\label{div-NS}
\left.
\begin{array}{c}
\displaystyle{ \nabla \cdot \textbf{u} =0 ~},\\
\displaystyle{ \rho\left(\frac{\partial \textbf{u}}{\partial t} + \textbf{u} \cdot\nabla\textbf{u}\right) =  - \nabla p + \nabla \cdot \mu ({\nabla \textbf{u}}+{\nabla \textbf{u}}^T) + \textbf{F}_s} ~,
\end{array}\right.
\end{equation} 
where $t$, $\textbf{u}$, $p$, and $\textbf{F}_s$ stand for time, velocity, pressure, and the surface tension force, respectively. The density and viscosity are expressed using the following formulation

\begin{equation}
\left.
\begin{array}{c}
\displaystyle{  \rho=\rho_{_{so}} + \left(\rho_{_w} - \rho_{_{so}}\right)  \mathcal{H}\left(\textbf{x},t\right) },\\
\displaystyle{  \mu=\mu_{_{so}} + \left(\mu_{_w}  - \mu_{_{so}}\right)  \mathcal{H}\left( \textbf{x},t \right)}.
\end{array}\right.
\end{equation}
\noindent
wherein $ \mathcal{H}\left( \textbf{x},t\right)$ represents a smoothed Heaviside function, which is zero in the dispersed phase (water) and unity in the stagnant phase (silicone-oil), while the subscripts `$w$' and `$so$' refer to the individual phases. The surface tension force $\textbf{F}_s$ is defined by using the hybrid formulation as in \cite{Shin_ijnmf_2009} and \cite{Shin_jmst_2017}
\begin{equation}
\textbf{F}_s=\sigma  \kappa_\mathcal{H} \nabla \mathcal{H},
\end{equation}
\noindent
where $\sigma$ refers to surface tension,  and $\kappa_\mathcal{H}$ is twice the mean interface curvature calculated from the Eulerian grid by using
\begin{equation}
\kappa_\mathcal{H} =  \frac{\boldsymbol{F}_L\cdot \mathbf{G}}{\sigma \mathbf{G}\cdot \mathbf{G},}
\end{equation}
where
\begin{equation}
\mathbf{F}_{L} =  \int_{\Gamma (t) } \sigma  \kappa \mathbf{n}  \delta   (\textbf{x}-\textbf{x}_f) ds,
\quad
\quad
\quad
\mathbf{G} =  \int_{\Gamma (t) }  \mathbf{n}  \delta   (\textbf{x}-\textbf{x}_f) ds.
\end{equation}
\noindent
Here, $\textbf{x}_f$ is the parametrisation of the interface, ${\Gamma (t) }$, and $\delta(\textbf{x}-\textbf{x}_f)$ is a Dirac delta function which vanishes everywhere except at the interface; $\textbf{n}$ is the  outward-pointing unit normal vector to the interface, and $ds$ is the length of the surface element. 

The incompressible Navier-Stokes equations (\ref{div-NS}) are solved by using  a second-order finite-difference method on a staggered grid \citep{Harlow_pf_1965,temam}. The computational domain is then discretised by a fixed, regular, Eulerian grid, and the spatial derivatives are approximated by standard centred difference discretisation, except for the non-linear term, which makes use of a second-order essentially non-oscillatory (ENO) scheme \citep{Sussman_1994}. As for the viscous term,  we use a second-order centred difference scheme. We have used the projection method to handle the incompressibility condition combined with a multigrid iterative method for solving the elliptic pressure Poisson equation \citep{Chorin_mc_1968,kwak_2003}. The numerical method uses an additional adaptive Lagrangian grid based on a hybrid front-tracking/level-set method \citep{Shin_jmst_2007,Shin_jmst_2017} to track the interface location. The geometrical information, the pressure and velocity variables are exchanged between the adaptive Lagrangian mesh and the fixed Eulerian grid following the Immersed Boundary Method of \cite{Peskin_jcp_1977}. The physical elements that form the Lagrangian interface are advected according to $\rm{d}\mathbf{x}_f/ \rm{d}t =\mathbf{V}$, where $\mathbf{V}$ is the interface velocity interpolated numerically from the fixed Eulerian grid, using a second-order Runge-Kutta method. Finally, the numerical method uses a domain-decomposition technique for its parallelisation and Message Passing Interface (MPI) for exchanging information between adjacent subdomains. More information regarding the full implementation of the numerical method can be found in \citet{Shin_jmst_2017,Shin_jcp_2018}.

\subsection{Numerical configuration and physical parameters}

\begin{table}
\caption{\label{tab:properties} Density and viscosity of the fluids used throughout this work \citep{Ibarra_2017}.}
\begin{center}
\begin{tabular}{cccc}
\hline
$\rho_{_w}$ & $\rho_{_{so}}$ &   $\mu_{_w}$ &  $\mu_{_{so}}$  \\ 
kg/m$^3$    & kg/m$^3$   &   $mPa. s $& $ mPa. s$       \\
998   &  824   &   1.0    & 5.4    
\end{tabular}
\end{center}
\end{table}

Figure \ref{fig:introductory_image} shows the three-dimensional computational domain, which is a rectangular box of size $20D \times 4D \times 4D$, where $D$ stands for the inner diameter of the nozzle (e.g., 4mm). The jet is produced when water leaves the cylindrical nozzle to enter progressively into the stagnant silicone oil. The physical properties of the fluids are given in table \ref{tab:properties}. An inflow boundary condition is applied to the nozzle on the left of the domain, i.e., at $(x = 0)$, which follows a simplified power-law turbulent velocity profile:
\begin{equation}\label{in_vel}
u\left(r,t\right) = \frac{15}{14} {U} \left(1-\left(\frac{r}{D/2}\right)^{28} \right)\left(1+A \sin\left(2\pi f t\right) \right).
\end{equation}

Here, ${U}$, $A$ and $f$, stand for the average injection velocity, amplitude and frequency, respectively, of the external pulsatile perturbation. The radial distance, $r$, within the jet measured from its centreline $(y_0, z_0)$ is $r=\sqrt{\left(y-y_0\right)^2 + \left(z-z_0\right)^2}$. The values for  $A$ and $f$ are informed by the previous work of  \cite {Ling_ijmf_2015} (e.g., $A=0.05$ m/s and $f=20$ Hz), and the same values have been used throughout the entire paper.

The numerical setup closely follows other computational studies; for example,  we impose a free boundary condition on the walls of the computational domain to let the fluid freely enter or leave the boundaries \citep{Taub_2013,Ling_ijmf_2015,Ling_jfm_2019}. A pressure outflow boundary condition is applied on the right surface of the domain to allow the fluid to exit the domain. The solid nozzle is treated as a no-slip surface \citep{asadi_asgharzadeh_borazjani_2018}.

The distance $\textbf{x}$, velocity $\textbf{u}$, time $t$, and pressure $p$ in equation (\ref{div-NS}) are rendered dimensionless using the following characteristic scales, $D$, $U$, $D/{U}$, and $\rho_{w} U^2$, respectively. Hence, the dimensionless control parameters governing the phenomena we will study are given by:

\begin{equation}\label{non_dim}
Re =\frac{\rho_wU D}{\mu_w} , \quad We =\frac{\rho_w U^2 D}{\sigma}, 
\end{equation}
where $Re$ stands for the Reynolds number (i.e. the ratio of inertial to viscous forces), and $We$ represents the Weber number (i.e. the ratio of inertial to capillary forces).  All the variables appearing in the equations and boundary conditions are rendered dimensionless using the aforementioned scalings, unless stated otherwise.
 
The first part of the results section corresponds to a discussion of  a regime map of jet dynamics in the $Re-We$ space. In this instance, we have used the M2-mesh (see table \ref{tab:mesh} in the Appendix) to perform the fully three-dimensional simulations. The selection of this mesh is dictated by the need to map out parameter space relatively rapidly before focusing on elucidating the details of the dynamics using the M3-mesh, which provides higher resolution, for four cases; the $Re-We$ combinations for these cases are listed in
table \ref{tab:cases}.  Information regarding the mesh-refinement study, resolution considerations, and the validation for this work are detailed in the Appendix.
\begin{table}
\caption{\label{tab:cases} The Reynolds-Weber number combinations for the four cases studied in detail with the  
M3-mesh (see table \ref{tab:mesh} in the Appendix).}
\begin{center}
\begin{tabular}{cccc}
\hline
Case   &  $Re$  	&  $We$     \\
1             & 1000     	&     7 \\	
2             & 1000    	&     100       \\
3             & 3260    	&     75       \\
4             & 6530       &    303      \\
\end{tabular}
\end{center}
\end{table}

\section{Results}\label{sec:Results}

\subsection{Interfacial dynamics: phase diagram in $Re-We$ space}\label{sec:parametric}

\begin{figure}   
 \includegraphics[width=0.92\linewidth]{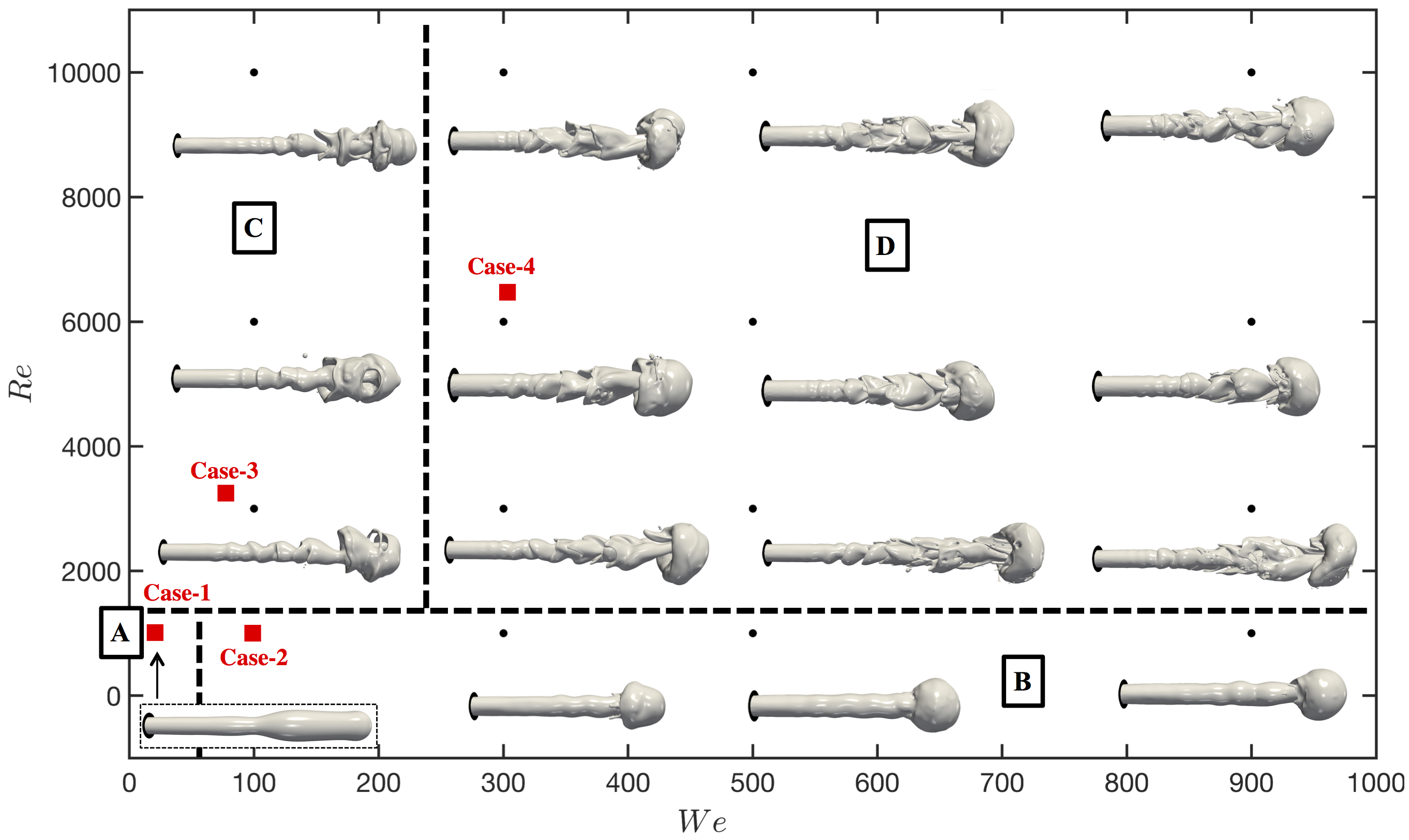}
 \caption{\label{regime_map} Regime map of the phenomenological interfacial dynamics in the $Re$-$We$ space using the $M2$-mesh (see table \ref{tab:mesh} in Appendix for mesh details). Four different regimes and their boundaries are identified. A snapshot of the flow corresponding to the three-dimensional representation of the interface for each simulated point (i.e., black marker) is shown. The red squares refer to the cases presented in table \ref{tab:cases}.}
 \end{figure}

 \begin{figure}
 \begin{center}
 \begin{tabular}{c}
 \includegraphics[width=1\linewidth]{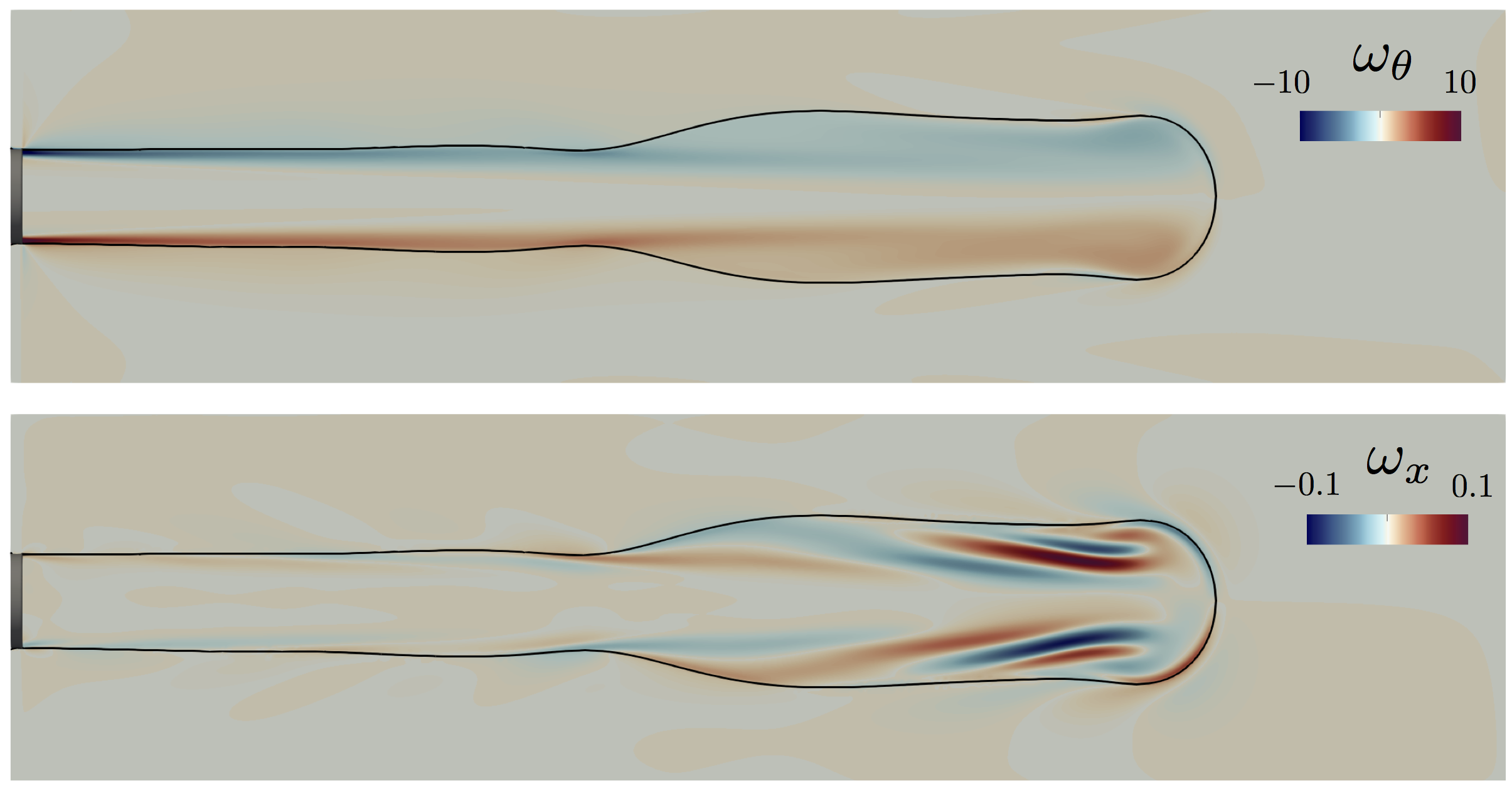}
 \end{tabular}
  \end{center}
 \caption{\label{case_1} Two-dimensional representation of the interfacial  location together with  $\omega_{\theta}$ (top panel) and $\omega_{x}$ (bottom panel) in the  $x-z$ plane  ($y=2$) for Case 1 in table \ref{tab:cases} ($Re=1000$ and $We=7$) at $t=24.38$. The colour represents the respective vorticity field, where appropriate scales are shown in each panel.}
 \end{figure}

 \begin{figure}
 \begin{center}
 \begin{tabular}{c}
 \includegraphics[width=1\linewidth]{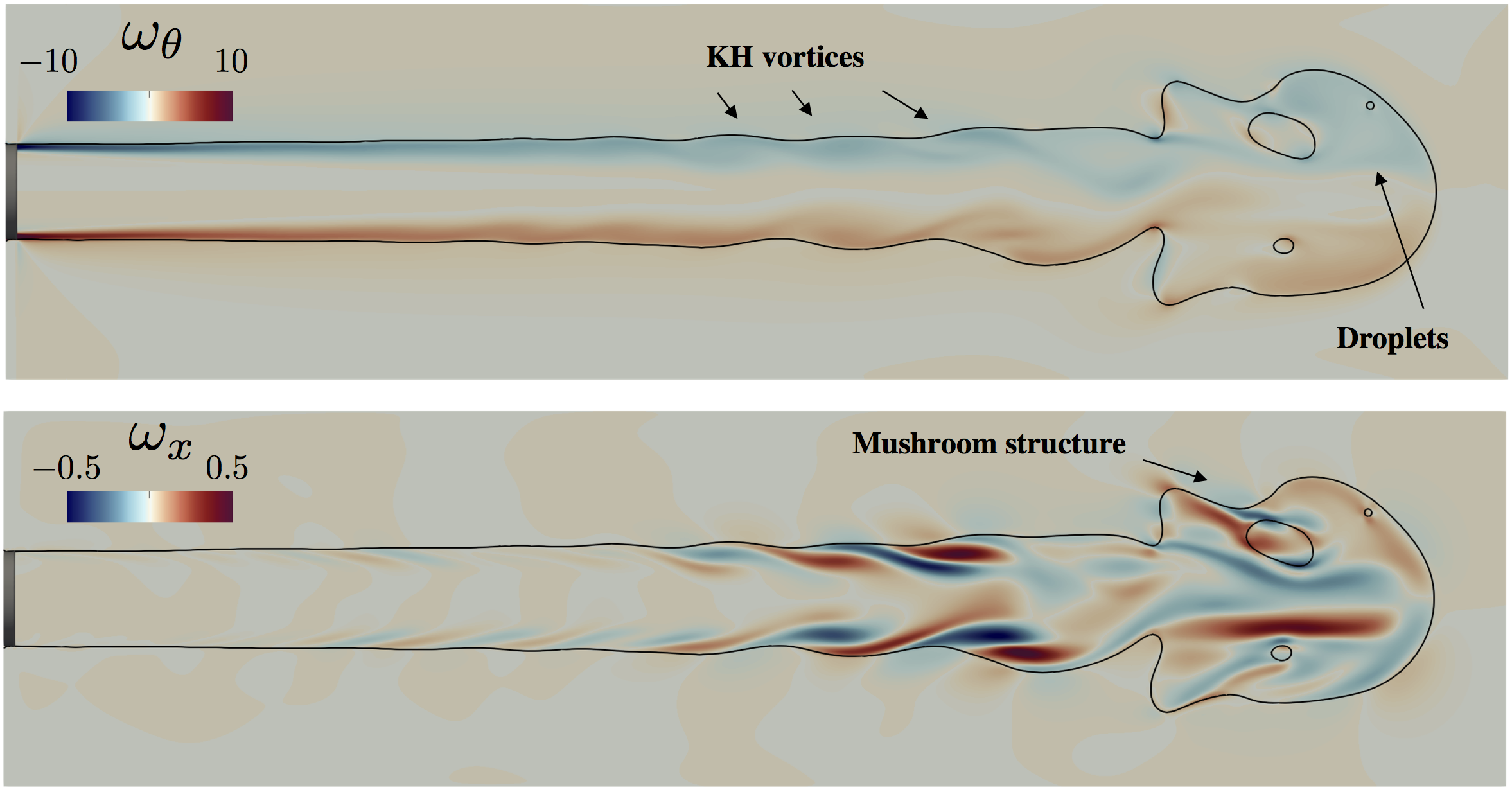}
 \end{tabular}
  \end{center}
 \caption{\label{case_2}  Two-dimensional representation of the interfacial location together with  $\omega_{\theta}$ (top panel) and $\omega_{x}$ (bottom panel) in the  $x-z$ plane  ($y=2$) for Case 2 in table \ref{tab:cases} ($Re=1000$ and $We=100$) at $t=26.87$. The colour represents the respective vorticity fields, where appropriate scales are shown in each panel.}
 \end{figure}
 
 \begin{figure}
 \begin{center}
 \begin{tabular}{c}
 \includegraphics[width=1\linewidth]{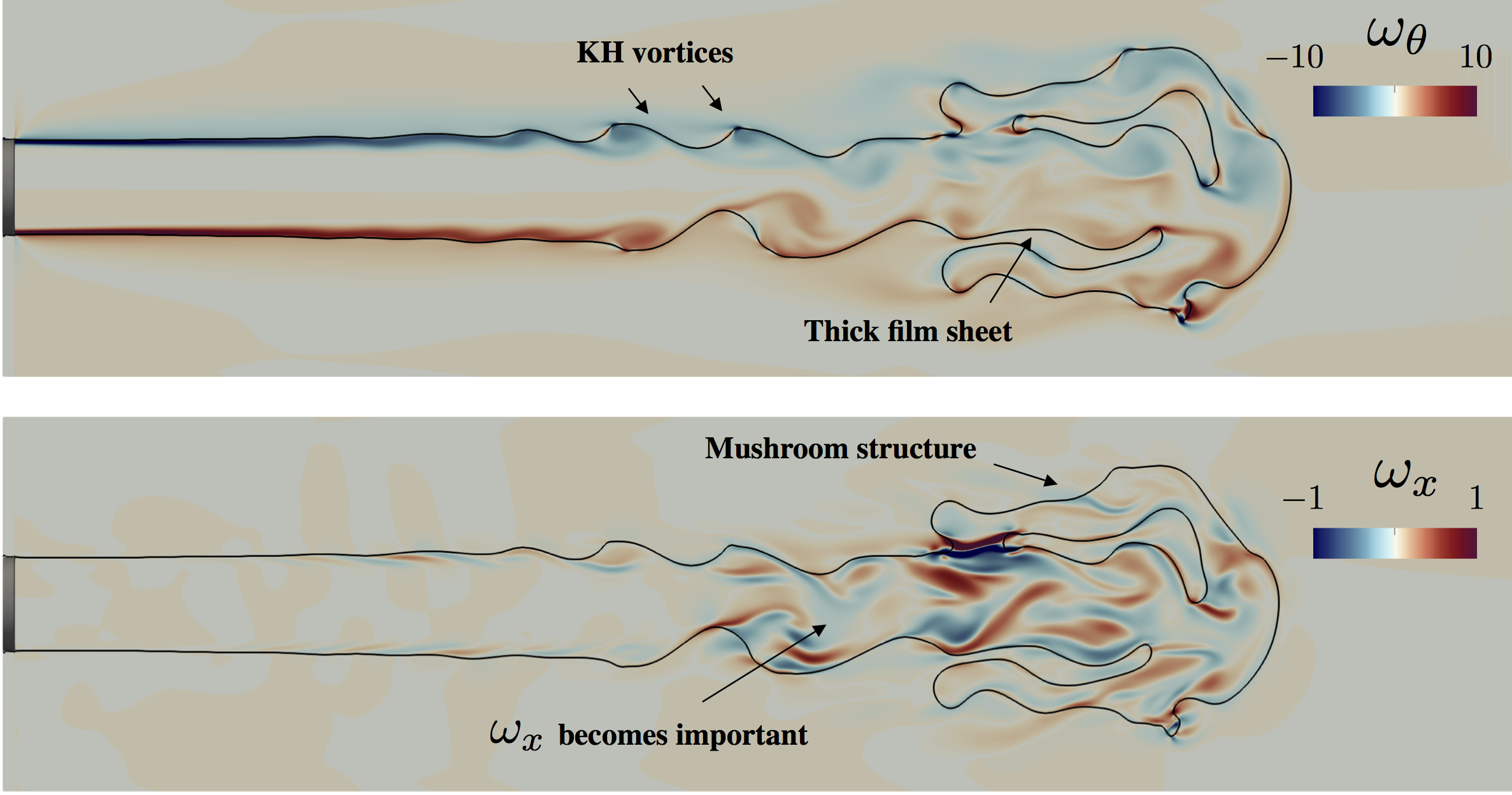}
 \end{tabular}
  \end{center}
 \caption{\label{case_3}  Two-dimensional representation of the interfacial location together with  $\omega_{\theta}$ (top panel) and $\omega_{x}$ (bottom panel) in the $x-z$ plane  ($y=2$) for Case 3 in table \ref{tab:cases} ($Re=3260$ and $We=75$) at $t=24.65$. The colour represents the respective vorticity fields, where appropriate scales are shown in each panel.}
 \end{figure}

We start the discussion of the results by  presenting  a phenomenological picture  of the interfacial dynamics in a phase diagram in  $Re-We$ space. The Reynolds and Weber numbers range between $10^3-10^4$ and $7-900$, respectively. Figure \ref{regime_map} shows the regime map in terms of the interfacial dynamics predicted from our numerical simulations. Several features emerge from this figure, which we have divided into four phenomenological regions based on the appearance of the interfacial structures. We aim to quantify the different jet behaviours by close inspection of the interplay between the vorticity field, $ \mathbf{\omega} =\nabla   
\times \textbf{u}$,  and the interface. Such in-depth analysis is carried out following the introduction of each region, utilising the higher resolution M3-mesh examples, as shown in table \ref{tab:cases}.

Region `A' in figure \ref{regime_map} is characterised by low Reynolds and Weber numbers and defined by the dominance of the capillary over inertial forces,  yielding an axisymmetric behaviour of the jet. For small $We$, the dominance of the surface tension forces results in the entrapment of the initial vortex head inside the discharged liquid-edge.  Then, the formation of an interfacial leading mushroom-like structure is not  observed. Figure \ref{case_1} shows the interplay between the azimuthal and streamwise vorticity components, $\omega_{\theta}$ and $\omega_x$, respectively for Case 1 (see table \ref{tab:cases}). We observe that $\omega_{\theta}$ exceeds $\omega_{x} $ by two orders of magnitude for the entirety of the jet, which explains the lack of deformation of the jet-core, and its axisymmetric shape. 

Region `B' is defined by low Reynolds and high Weber numbers. As shown in figure \ref{regime_map}, the snapshots of the interface in this region of parameter space reveal the development of interfacial waves as well as the formation of a `mushroom'-like structure at the jet leading edge. This is due to the rolling of the boundary layer at the edge of the solid nozzle as a result of the initial vortex ring  (more details in Section 3.2 ). Figure \ref{case_2} shows that although the magnitude of $\omega_{\theta}$ exceeds that of  $\omega_x$  by almost two orders of magnitude for Case 2, the relative significance of $\omega_x$ has increased in comparison with Case 1. This is correlated with the corrugations which are observed on the main body of the jet in Case 2 that appear to be largely absent in Case 1. These corrugations are related to the development of Kelvin-Helmholtz (KH) instabilities that arise from the velocity difference between the injected jet and the surrounding, initially stagnant phase. As shown in figure \ref{case_2}, the mushroom structure is also accompanied by the formation of droplets within it. More information regarding the mechanisms which induce droplet formation is provided in Section \ref{hole_section}.

Region `C' is characterised by low Weber and high Reynolds numbers. To elucidate the mechanisms at play in this region, we refer to Case 3 (see table \ref{tab:cases}), as shown in figure \ref{case_3}. It is seen clearly that the magnitude of $\omega_{\theta}$ is an order of magnitude larger than that of  
$\omega_{x}$. The higher levels of inertia in Case 3, in comparison to Case 2, leads to formation of pronounced KH instabilities with vortical rings close to the interface, yielding spanwise interfacial deformations. Close inspection of figure \ref{case_3} reveals that the mushroom structure, which is also present in Case 3, undergoes significant deformation leading to the formation of a toroidal sheet, which envelopes that main body of the jet. 

In region `D', the flow is characterised by both high Reynolds and Weber numbers. Here, it is seen from the snapshots of the interface in figure \ref{regime_map} that the interfacial dynamics in this region of $Re-We$ space are the most complex. The mushroom structures suffer severe deformation as does the main body of the jet which also features the formation of lobes arising from the KH-induced corrugations. It is also evident that flow in region D is also accompanied by droplet generation. In Section \ref{case_4_5} below, we focus on Case 4 in figure \ref{regime_map} and provide an extensive explanation of the mechanisms linking droplet formation to vortex-interface interaction.   

With the purpose of providing a better understanding of the vortex-surface interaction for the development of the three-dimensional instabilities, we have analysed the rate of change of the vorticity production by taking the curl of the momentum equation (i.e., equation (\ref{div-NS})), which can be written as

\begin{equation}
    \frac{D\boldsymbol{\omega}}{D x}=\left( \boldsymbol{\omega} \cdot \nabla \right )\mathbf{u} + \nabla \times\left ( \frac{\nabla \cdot \mathbf{ \tau }}{\rho}  \right ) +\frac{1}{\rho^2} \nabla \rho \times \nabla P + \nabla \times \left (\frac{\mathbf{F}_s}{\rho}\right) 
\end{equation}

\noindent
where $\mathbf{\omega}$ stands for the vorticity and   $\mathbf{ \tau }$ represents the viscous stress tensor. The right-hand-side represents the vortex stretching, the vorticity generation as a result of the  viscous diffusion, the  effect of density variation (`baroclinic torque') and the surface tension forces, respectively.
\citet{Jarrahbashi_jfm_2016} have shown that the baroclinic term is responsible for the 3D instabilities for large density ratios $O(10^{-2})$, meanwhile,  at lower density ratios $O(10^{-1})$, the streamwise vorticity generation is dominated by the azimuthal tilting and radial tilting of the vortex rings. For our study, the contribution  of the baroclinic term to the vorticity generation is negligible as  the density ratio is of the same order of magnitude (i.e., $\sim O(10^0)$).

Table \ref{table_vorticity} collects the results of a simple dimensional analysis to study the dominant terms of the vorticity generation equation for the  selected cases shown in Table \ref{tab:cases}. On this basis, the  viscous diffusion term scales as $\mu_w U/\rho\Delta x^3$, and the surface tension term scales as
$\sigma \kappa / \rho\Delta x^2$ (similar to  what was presented by \cite{zandian_sirignano_hussain_2019,Jarrahbashi_jfm_2016}). Finally, the vorticity stretching term $\left ( \boldsymbol{\omega} \cdot \nabla \right ) \textbf{u}$ has been computed numerically. For Regions A-C, the vortex stretching term is negligible in comparison to the other terms, which  explains  the nondevelopment of the 3D  interfacial instabilities on the surface of the jet core. Those regions are characterised by a balance/competition between the surface-tension and viscous terms giving rise to the different interfacial instabilities explained above. Finally, in the  inertia-dominated region, the three terms on the right-hand-side of the vortex generation equation are balanced, and their competition  determines the interfacial topology. Thus, the vortex stretching term from the vorticity equation plays the major role in the development of the three-dimensional interfacial destabilisation  for turbulent jets with density ratios of the same order of magnitude. Similar conclusions were drawn by \citet{Jarrahbashi_jfm_2016}.  

\begin{table}
    \centering
    \caption{Scalings for the terms of the vorticity transport equation (nondimensional values).
    \label{table_vorticity}}
    \begin{tabular}{cccc}
      Cases   & Vortex stretching & Viscous & Surface tension \\
       & $\left ( \boldsymbol{\omega} \cdot \nabla \right ) \textbf{u}$ & $\nabla \times\left ( \dfrac{\nabla \cdot \mathbf{ \tau }}{\rho}  \right) $ & $\nabla \times \left (\dfrac{\mathbf{F}_s}{\rho}\right ) $ \\      
      \hline
       1  & $\sim O(10^0)$ &  $\sim O(10^5)$ &  $\sim O(10^5)$\\
       2  & $\sim O(10^1)$ &  $\sim O(10^5)$ &  $\sim O(10^4)$\\
       3  & $\sim O(10^3)$ &  $\sim O(10^5)$ &  $\sim O(10^5)$\\
       4  & $\sim O(10^5)$ &  $\sim O(10^6)$ &  $\sim O(10^5)$\\
    \end{tabular}
\end{table}

\subsection{Interfacial dynamics explained through vortex-surface interaction \label{case_4_5}}

\begin{figure}
\begin{center}
 \includegraphics[width=0.8\linewidth]{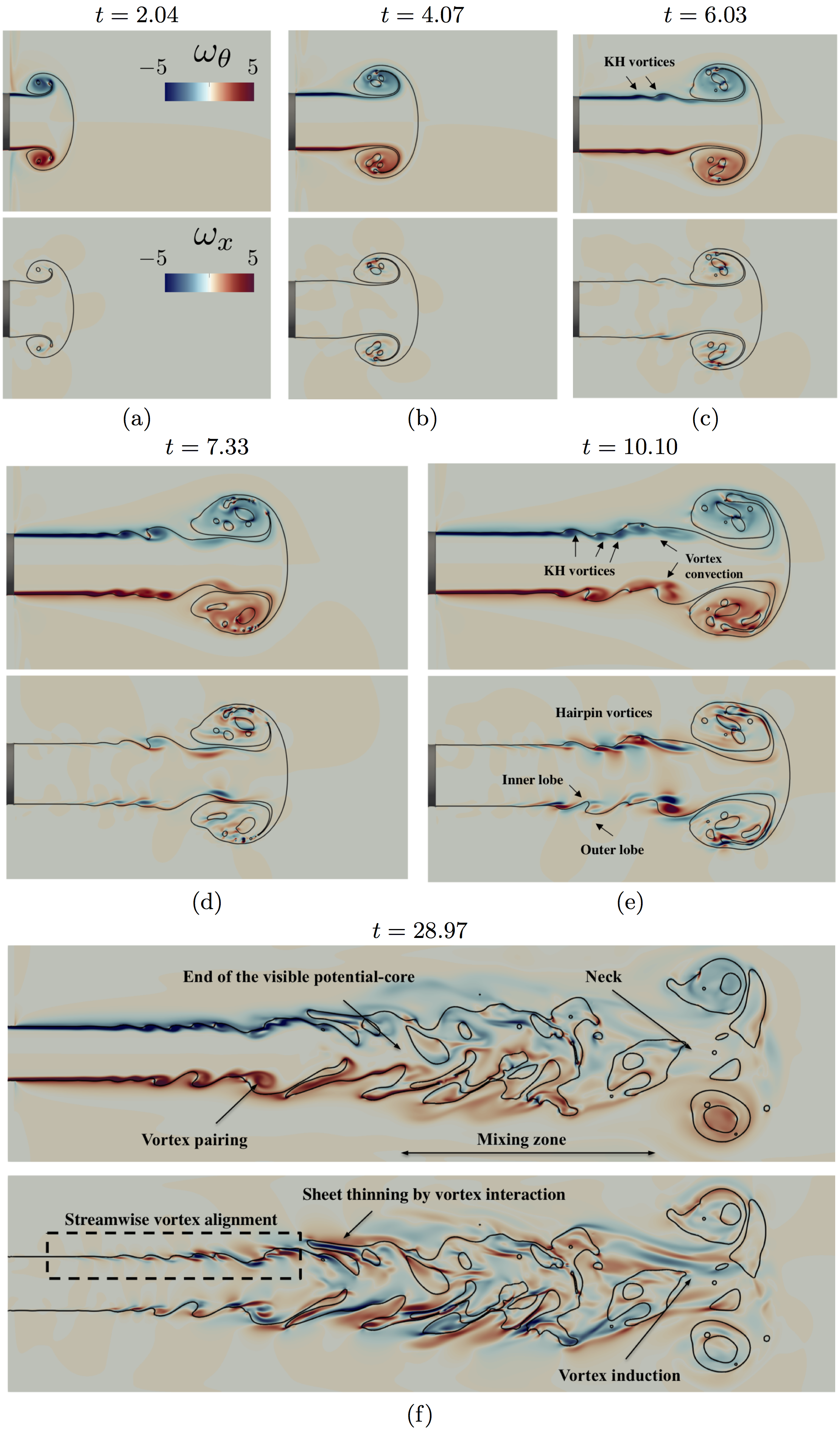}
\end{center}
\caption{\label{figure_wx_wy}  Two-dimensional representation of the spatio-temporal evolution of the interface for Case 4 in table \ref{tab:cases}  ($Re=6530$ and $We=303$) together with $\omega_{\theta}$ (top panel of each sub-figure) and $\omega_{x}$ (bottom panel of each sub-figure) in the  $x-z$ plane  ($y=2$)  for the dimensionless times shown in each panel. The colour represents the respective vorticity fields, where appropriate scales are shown in each panel.} 
\end{figure}

This section focuses on the inertia-controlled Region D and Case 4 as a principal example of the flow in this region.  In figure \ref{figure_wx_wy}, we examine the spatio-temporal interfacial dynamics overlaid with the magnitude of $\omega_\theta$ and $\omega_x$  in the $x-z$ plane.  During the early stages of the injection, a vortex ring is initially formed by the rolling of the boundary layer at the edge of the solid surface. The physical mechanism which results in the formation of the leading vortex ring is in agreement with previous work, such as \citet{gharib_rambod_shariff_1998,marugan_2013,asadi_asgharzadeh_borazjani_2018}. Once the jet enters the domain, the rotation of the head vortex drives the entrainment of stagnant phase as it is transported radially outwards. The initial vortex has a profound effect on the interfacial dynamics with the formation of a ``mushroom-like?  structure. As time evolves, the entrainment and formation of a toroidal liquid sheet of  stagnant phase are observed inside  the mushroom structure (see for example figure \ref{figure_wx_wy}a,b). Upstream of the mushroom, it is seen that the KH instability develops (see figure \ref{figure_wx_wy}c) amplified by the  pulsatile-injection into waves which cause a local adverse pressure gradient by virtue of the local interfacial curvature (see figure \ref{figure_wx_wy}d). With increasing time, we observe the formation of outer lobes as a result of the entrainment of the stagnant phase in the jet-core (see figure \ref{figure_wx_wy}e).

Figure \ref{figure_wx_wy} also  highlights the spatio-temporal development of the azimuthal and streamwise components of the vorticity.  At early times, vorticity generation coincides with the velocity boundary layer attached to the interface, which corresponds to strong tangential flow near the interface. The  roll-up of the shear layers, and subsequently, the vortex roll-up gives rise to the formation of KH vortex rings close to the interface (see figure \ref{figure_wx_wy}c). As time increases, 
the vorticity boundary layer is convected towards the jet-core  (see figure \ref{figure_wx_wy}e).  Additionally,  vorticity dissipates strongly in the stagnant phase due to the damping effect of the viscosity. 

\begin{figure}
\begin{center}
 \includegraphics[width=1.0\linewidth]{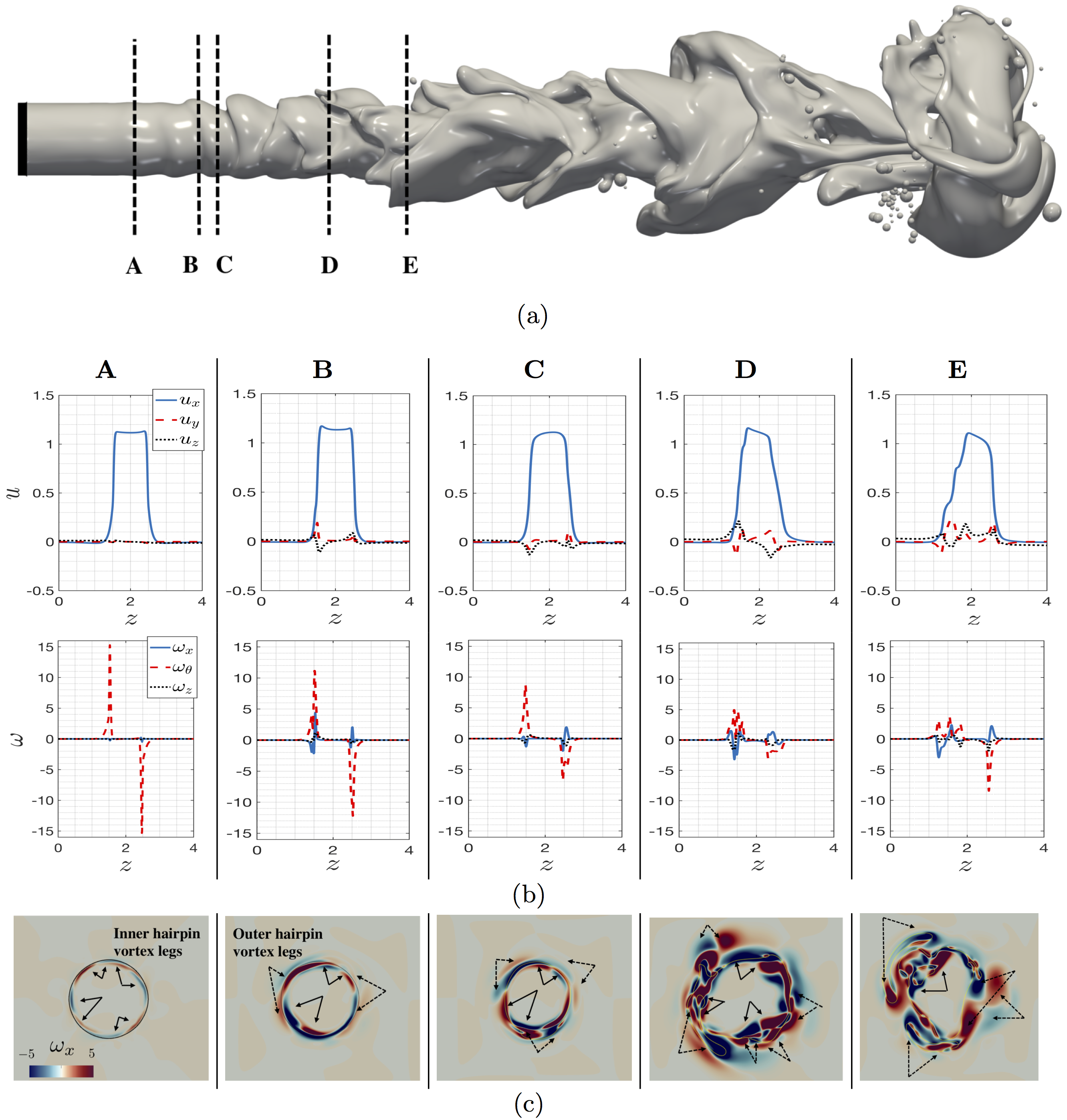}
 \end{center}
 \caption{\label{wx_cross_section_panels}
 Three-dimensional representation of the interface for Case 4 ($Re=6530$ and $We=303$) at $t=28.97$, showing the spatial locations `A'-`E', (a). The streamwise location of the `A'-`E' probe lines correspond to $x=(1.50, 2.50, 2.75, 4.25, 5.25)$, respectively. (b) Velocity (top panel) and vorticity profiles (middle panel) in the $y=2$  plane for each probe location. (c) Streamwise vorticity in the $y-z$ plane for each sampling location. The arrows show examples of identified hairpin-vortex legs. Solid and dashed arrows correspond to inner and outer hairpin vortex legs, respectively. The colour represents the streamwise vorticity field, $\omega_x$.} 
\end{figure}

\begin{figure}   
\centering
 \includegraphics[width=1.0\linewidth]{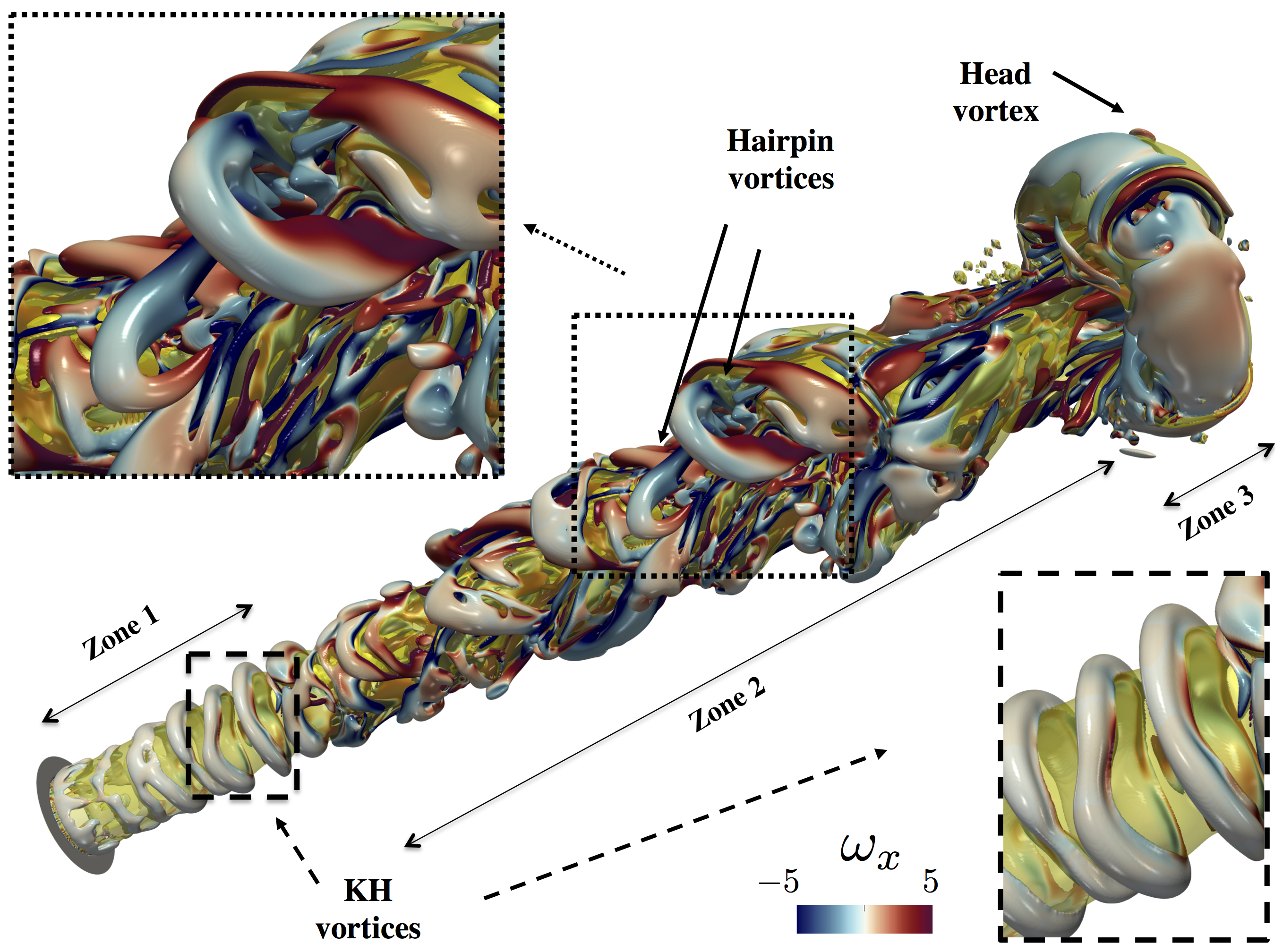}
 \caption{\label{3d_vorticity} Illustration of the coherent vortical structures close to the interface for Case 4 ($Re=6530$ and $We=303$) at $t=28.97$. The coherent vortical structures are visualised by the Q-criterion with a value of $Q=0.1$, where the colour represents the streamwise vorticity field, $\omega_w$. Magnified views of the KH-vortex rings near the injection point, and the hairpin-vortical structures in the jet core are also shown.}
\end{figure}

Attention is now turned towards the competition between $\omega_{\theta}$ and $\omega_{x}$.  In the region adjacent to the injection point, the azimuthal vorticity component dominates over its streamwise counterpart by two orders of magnitude; this dominance  is reflected by the fact that the KH vortex rings shown in figure \ref{figure_wx_wy}c are essentially quasi-axisymmetric.  As the flow develops downstream (see figure \ref{figure_wx_wy}d,e), $\omega_{x}$ becomes comparable in magnitude to $\omega_{\theta}$ leading to  streamwise stretching of the KH vortex rings to form hairpin-shaped vortices. Hairpin vortical  or `horseshoe' structures were proposed by \citet{Theodorsen1955} to describe the features of turbulence dynamics related to the existence of a shear or boundary layer near a wall. A hairpin vortex is made from a `vortex-head' which is the arched region farther away from the boundary layer. The head is connected to the free surface or wall by its `vortex-legs'. The hairpin-head orientation results from a balance between the effect of shear flow and the local velocity. 

Interestingly, the streamwise alignment of  hairpin vortices  near the surface (see figure \ref{figure_wx_wy}e,f) has been reported previously by \citet{Jarrahbashi_jfm_2016} and \citet{Zandian_jfm_2018} for coaxial round and planar jets, respectively. In spite of the absence of a coaxial phase in our case, we observe the same phenomenon. Additionally, we observe the successive alignment of opposite-signed vortices, where the angle between each vortex pair is determined by the induction effect of the opposite-signed `neighbour' vortex. Similar arrangement of vortices has been reported previously by \citet{jeong_jfm_1997} for boundary layer turbulence, and \citet{davoust_jfm_2012} for compressible homogeneous jets. Further details about the specific alignment of the vortical structure is provided below.  

Figure \ref{wx_cross_section_panels}a presents a three-dimensional representation of the jet at $t=28.97$ alongside five spatial locations which will be used to  show the evolution of vorticity and velocity across the jet core. The specific locations were chosen as follows: panel `A' is representative of the dynamics near the injection point, where there is little disturbance to the interface;  panel `B' displays the cross-section of the jet in a jet-ring (i.e., the part of the interface where we observe the KH-waves crest), whereas panel `C' depicts a jet-braid (i.e., the KH-waves  trough); panels `D' and `E' portray the jet core dynamics where lobe formation is present. Figure \ref{wx_cross_section_panels}b shows the velocity and vorticity line profiles at the $y=2$ plane alongside the streamwise vorticity in the $y-z$ plane for each of those locations. Near the injection point (location `A'), $\omega_{\theta} >> \omega_{x}$, which is in agreement with the lack of deformation of the vortical structures in the streamwise direction (see also figure \ref{figure_wx_wy}f). The tangential motion of the fluid close to the surface results in the emergence of the velocity boundary layer. In turn, this causes the  formation of two large peaks of opposite signs for $\omega_{\theta}$. As the flow evolves downstream, $\omega_{\theta}$ still dominates the physics, although, its value has been reduced in favour of an increase in  $\omega_{x}$, and consequently the local formation of corrugations on the jet-core (location `B'). Additionally, the $\omega_{\theta}$ peaks have widened in their base due to the growth of the velocity boundary layer as the jet moves downstream \citep{Liepmann_jfm_1992}. Further downstream (location `D' and `E'), the interface has lost its cylindrical shape due to the entrainment of the stagnant phase as $\omega_{x}$ becomes comparable in magnitude to $\omega_{\theta}$. Several peaks can be seen in the vorticity profiles as the probe line passes through several lobes. By inspecting the instantaneous velocity-fields in locations `D' and `E', we observe that the main component of velocity is associated with the direction of injection, where there is a  decay of its mean centreline value as flow evolves downstream, which is in agreement with \citet{Pope_2000} for single-phase jets.  The velocity decay leads to a reduction in the velocity difference between the injected and stagnant phases, which attenuates the interfacial shearing, supporting the development of the vortical rings close to the surface of the jet. Notice that the $y$ and $z$ velocity-components are positive or negative close to the surface, showing the entrainment of injected fluid towards the stagnant viscous phase or vice versa. 

Next our attention is turned towards the physical mechanisms which lead to the alignment of the  hairpin-like vortical structures along the free-surface. Figure \ref{wx_cross_section_panels}c shows cross-section panels with $\omega_{x}$ of the jet at the streamwise locations shown in figure \ref{wx_cross_section_panels}a. 
As depicted in panel `A',  $\omega_{x}$ is mainly confined inside of the injected stream, where hairpin vortex legs are observed (see first panel of figure \ref{wx_cross_section_panels}c).  In panel `B', the outer layer of vorticity comes from the braid located immediately downstream, which starts to roll-up around the ring. Further analysis of panel `B' shows the existence of additional pairs of vortex legs which are $180^{\circ}$ out-of-phase with respect to the inner layer, and subsequently they correspond to a second hairpin vortex layer with opposite direction (e.g., upstream direction). Similarly, in panel `C' we observe  that the inner layer of the vorticity comes from the ring located upstream, whereas the outer layer comes from the ring located immediately downstream. Therefore, we can conclude that the alignment of the hairpin vortices is a result of the existence of a vortex-induction mechanism which causes the reorientation of  vortical structure within the ring and braid skeleton. This phenomenon is in agreement with \citet{Brancher_jfm_1994} for homogeneous jets, \citet{Jarrahbashi_jfm_2016} for the coaxial atomisation of a round liquid jet, \citet{Zandian_jfm_2018} for coaxial atomisation of planar jets, and \citet{bernal_jfm_1986} for plane mixing layers. The analysis between the panels `B' and `C', can be extended to other locations of the jet where the core has undergone further interfacial development. For example, in panels `D' and `E' we observe the same distribution of inner and outer layer of hairpin vortex legs. The alignment of vortical structures has a detrimental effect on the interfacial dynamics, which will be shown in section \ref{hole_section}.

Additionally, we have used the Q-criterion to present a three-dimensional visualisation of the vortices in the present study. The Q-criterion was described by \citet{Hunt_CTR_1988} as a quantity which measures the dominance of vorticity $\omega$ over strain $s$, $Q=1/2     (  \left \| \omega \right \|^2 - \left \| s \right \|^2 )$. Figure \ref{3d_vorticity} shows the  spatial development  of the coherent structures for $Q=0.1$. Near the injection point, quasi-axisymmetric KH vortex rings are located close to the  surface (a magnified view is shown). As explained previously, when $\omega_{x}$ and $\omega_{\theta}$ are of comparable magnitude, the KH vortices are stretched downstream or upstream. The topological shape of these vortical structures resembles the instantaneous hairpin-like vortical structures reported in experiments and numerical simulations by \citet{head_jfm_1981,zhou_jfm_1999,Zandian_jfm_2018}. Outer hairpin vortices (a magnified view is also shown) are observed clearly and the inner hairpin vortices not as clearly as they are localised underneath the interface. Further downstream, we observe  the `head-vortex' covering the mushroom-like structure. Inside of this structure, the vortices are unstable and break down. 

To conclude, we have identified three different zones in the transient flow field which are associated with  different vortical dynamics. We present the different regions at $t=28.97$ (see figure \ref{3d_vorticity}). Zone 1 starts from the injection point and extends up to the deformation of the free-surface ($x \sim 1.8$). This zone is characterised by the dominance of $\omega_{\theta}$. Zone 2, which extends from $x \sim 1.8$ up to $x \sim 11.5$ (e.g., behind the mushroom-like structure). This region is characterised by the interfacial deformation of the jet-core by KH vortex rings and their posterior deformation by the competition between $\omega_{x}$ and $\omega_{\theta}$.  As the flow moves downstream, $\omega_{x}$ becomes responsible for the entrainment of the stagnant phase to form interfacial lobes (this agrees with \cite{Liepmann_jfm_1992} for homogeneous jets). In this region, vortex rings pair up and merge together. Additionally,  the ending of the visible-potential core of the jet and the beginning of the mixing region is also observed. Zone 3, expands from the end of zone 2 up to the leading edge of the mushroom-like structure. A large vortex-ring dominates the dynamics in this region (so-called `cap-vortex', \cite{zandian_sirignano_hussain_2019}), wherein its interaction with upstream vortex structures via means of velocity induction leads to the formation of the neck, connecting the mushroom-like structure with the cylindrical body of the jet. The narrowing of the neck as a result of the streamwise convection of vortical structures is in agreement with the work of \cite{asadi_asgharzadeh_borazjani_2018}.

\subsection{Cascade mechanism for droplet formation: hole-formation genesis \label{hole_section}}

\begin{figure}
\centering
 \includegraphics[width=1.0\linewidth]{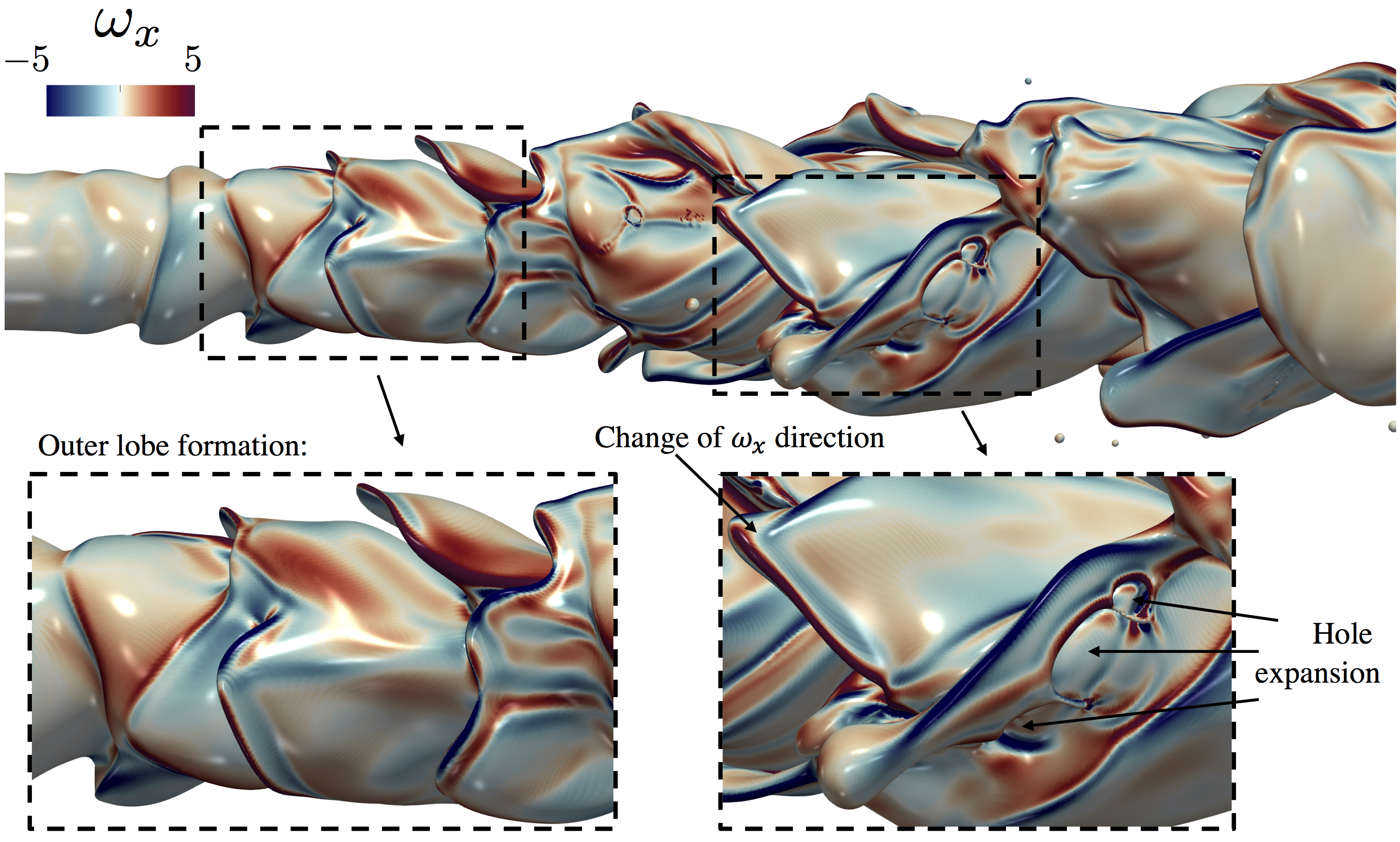}
\caption{\label{interface_coloured_vorticity} Three-dimensional representation of the surface of the jet for Case 4 ($Re=6530$ and $We=303$) at $t=28.97$ showing the spatial formation of outer lobes. The colour represents the streamwise vorticity field, $\omega_x$.}
\end{figure}

\begin{figure}   
\centering
 \includegraphics[width=1.0\linewidth]{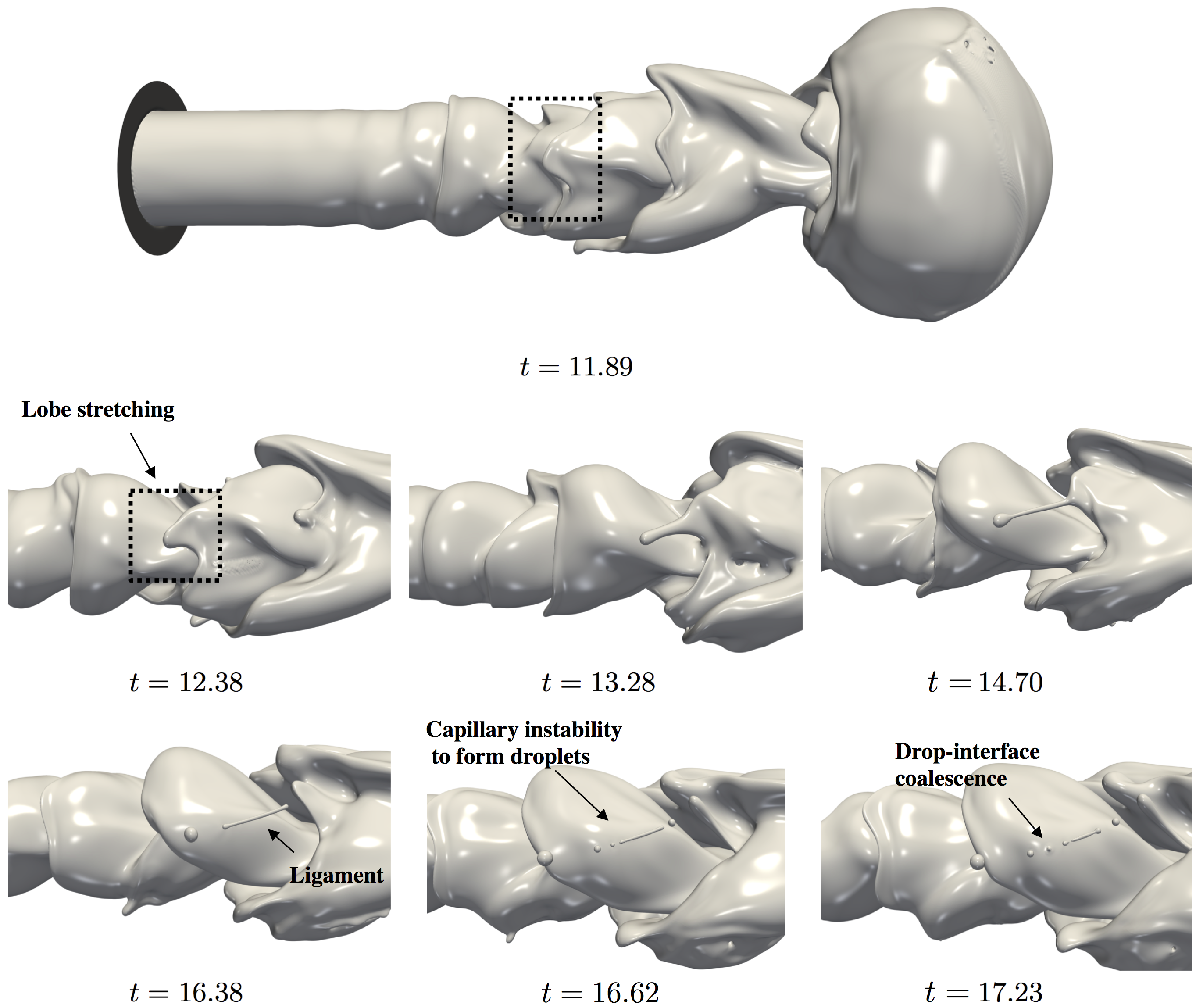}
\caption{\label{ligament_formation} Three-dimensional representation of the interface showing the spatio-temporal evolution of the formation of injected water-droplets via the stretching of the outer lobe for Case 4 ($Re=6530$ and $We=303$). Dimensionless times are shown in each panel.  } 
\end{figure} 

\begin{figure}
\centering
 \includegraphics[width=1.0\linewidth]{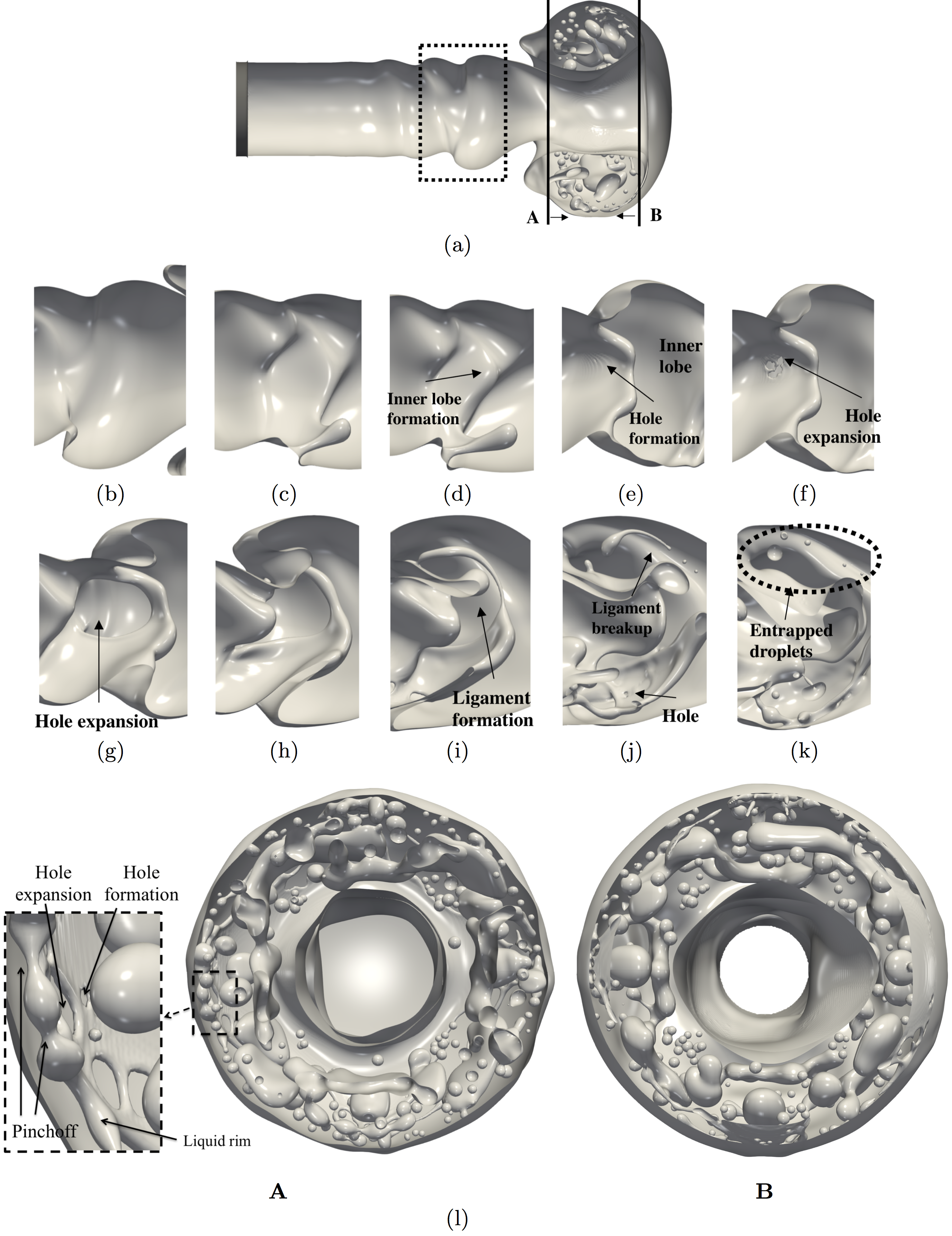}
\caption{\label{inner_lobe_mechanism}
(a) Three-dimensional representation of the interface 
for Case 4  ($Re=6530$ and $We=303$) at $t=7.33$ showing a selected region in the jet core where inner lobes will form, together with two probe locations in the leading-edge structure, where the arrows indicate the direction of the view. (b-k) Cascade mechanism for the formation of  entrapped oil-droplets within the jet core  at 
$t=(8.84, 9.96, 10.10, 10.92, 11.36, 11.41, 12.06, 13.61, 16.38)$, respectively. $(l)$ Illustration of the entrapped oil-droplets inside of the mushroom-like structure at $t=7.33$ from the back (probe `A') and front (probe `B')  of the structure, respectively. A magnified region of the structure is also shown to illustrate the hole-formation behind the rim-edge.} 
\end{figure}

\begin{figure}
\centering
 \includegraphics[width=1.0\linewidth]{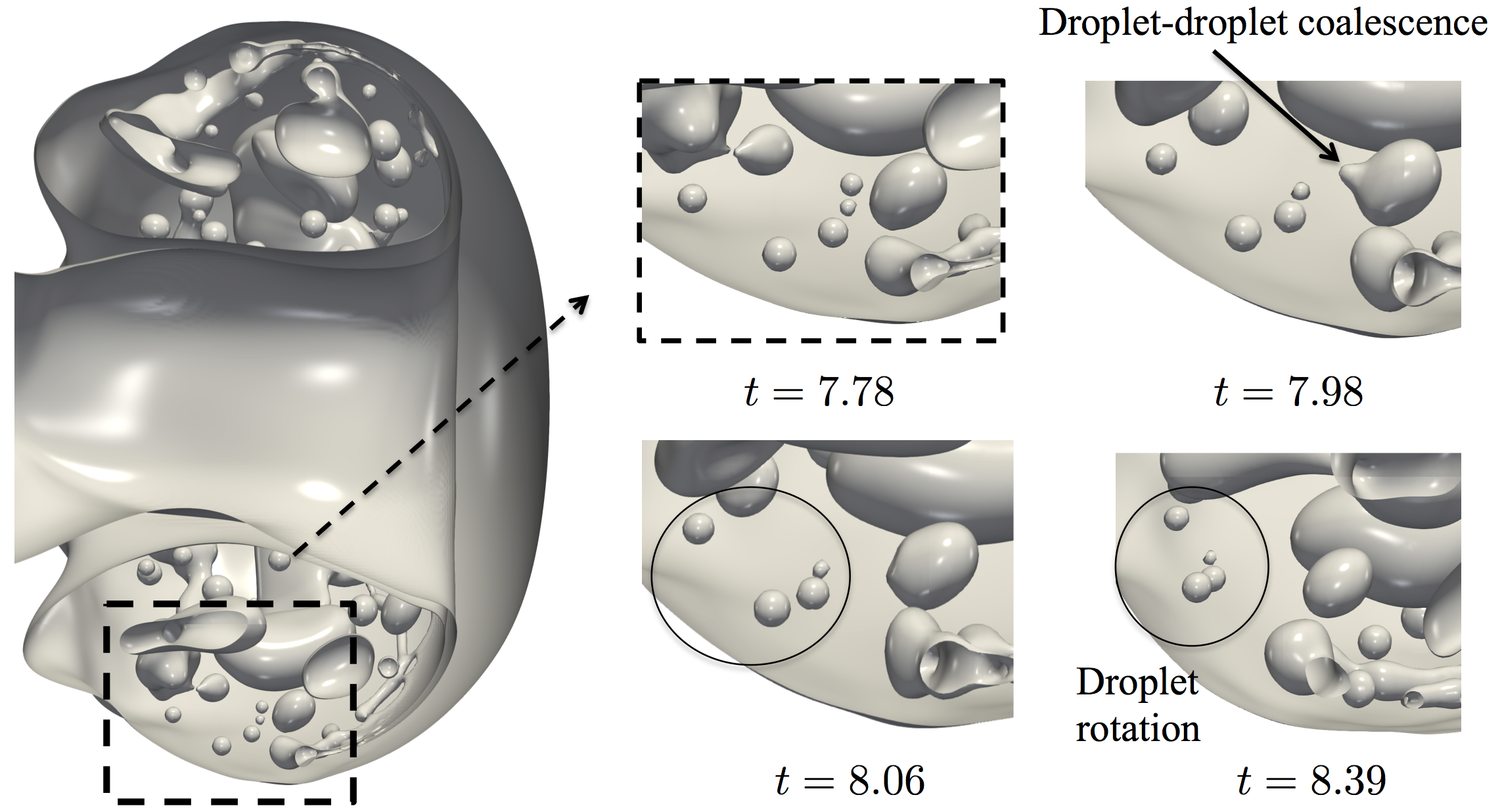}
\caption{\label{droplet_coalescence}
 Three-dimensional representation of the interface 
for Case 4  ($Re=6530$ and $We=303$) at dimensionless times shown in each panel. A magnified region of the structure is also shown to illustrate droplet-droplet coalescence and the droplet-rotation inside of the leading-edge structure.} 
\end{figure}

In the previous section, we discussed the vortex dynamics of the transient jet. This section focuses on the effect of vortices on the formation of inner/outer lobes, the thinning of which leads to genesis of droplets. Figure \ref{interface_coloured_vorticity} shows a close view of the interfacial jet dynamics where outer lobes are present. The formation of lobes is observed along the spanwise direction of the jet. Close inspection of the interface shows it has assumed the shape of the surrounding hairpin vortical structures, where a change of the vorticity direction is observed for subsequent lobes (similar observations have been made by \citet{Jarrahbashi_jfm_2016,Zandian_jfm_2018}). Consequently, the alignment of the hairpin vortices induces the thinning of the outer lobes, leading to formation of holes, which expand and eventually give rise to injected water-droplets.

Figure \ref{ligament_formation} shows the temporal stretching of an outer lobe to form a ligament, and eventually droplets. The ligament orientation is linked to the vortical structures, as suggested above. A bulbous-tip is formed  at the edge of the ligament driven by capillarity. The ligament detaches from the jet-core; then it retracts driven by capillarity to form droplets from both ends according to the so-called `end-pinching mechanism' \citep{notz_basaran_2004}. When the stagnant phase has enough momentum, the ligament is dragged into the stagnant phase, setting its  characteristic radial length (similarly to what has been reported by \cite{lasheras_villermaux_hopfinger_1998}). The ligament is characterised by having a small thickness prior to its detachment from the base. However, the droplets formed after the breakup process have a larger length scale (in agreement with \citet{Villermaux_arfm_2007}).

Figure \ref{inner_lobe_mechanism} shows the formation of droplets inside the jet core, and the mushroom structure. Specifically, figures \ref{inner_lobe_mechanism}b-k depict the formation and elongation of the inner lobe in the streamwise direction. As before, the stretching of the interface gives rise to a  thin sheet as a result of the mutual induction of neighbouring hairpin-vortices. The hairpin vortex located above the liquid sheet induces flow downwards, whereas the vortex located under the sheet induces flow upwards. Under the joint action of the vortical structures, the liquid sheet thins until it is perforated (see figure \ref{inner_lobe_mechanism}e). Following the formation of the hole, retraction of the liquid sheet is radially driven by capillarity (see figures \ref{inner_lobe_mechanism}f and \ref{inner_lobe_mechanism}g), and as shown in figure \ref{inner_lobe_mechanism}i, it subsequently gives rise to the formation of a curved liquid thread or ligament. Once more, ligament retraction is driven by capillarity where bulbous ends that form initially undergo `end-pinching' to form smaller droplets, as shown in figures \ref{inner_lobe_mechanism}j-k. The inertia-induced mechanism for the sheet-thinning presented here has been previously reported by \citet{Jarrahbashi_jfm_2016}.

The analysis will now focus on the dynamics underneath the mushroom-like structure. Figure \ref{inner_lobe_mechanism} shows entrapped oil-droplets inside of the leading-edge structure. The formation of these droplets is also linked to the interfacial rupture of the toroidal liquid sheet. The liquid sheet retracts driven by surface tension to accumulate liquid in a rim at its edge. The rim experiences a spanwise destabilisation (i.e., Rayleigh-Plateau type), which leads to a non-uniform rim-radius. The film retraction is also accompanied by the formation of interfacial capillary waves that precede the rim. The capillary waves vary the film thickness, and consequently induce the perforation of the film adjacent to the rim in regions where the film thickness is sufficiently small (in agreement with \citet{Mirjalili_snh_2018}). The radial expansion of multiple holes yields the formation of liquid threads which experience a capillary instability  to produce droplets.
Following their formation, the entrapped oil-droplets rotate inside the leading structure due to their interaction with the vorticity field (not shown here). The rotation leads to complex interfacial phenomena such as coalescence or collision (similar to that reported by \citet{Desjardins_as_2010}). The coalescence has been observed not only between droplets, but also between droplets and ligaments or droplets and the jet core (see figure \ref{droplet_coalescence}).

Next, we aim to explain the validity  of the hole formation mechanism as a result of liquid film thinning. Our numerical method does not consider the destabilising effect brought about by the van der Waals attractive forces  as the film thickness tends to zero.  \cite{Churaev_springer_1987} reported that while the thinning takes place, the dynamics of the film enters into an asymptotic behaviour leading to its rupture, depending only on a balance between viscous, van der Waals, and surface tension forces. After the film puncture, the circular hole expansion is driven by  capillarity, and fluid is accumulated in a circular rim which grows in size as it moves away from the puncture point. The retraction speed, $V_{TC}$ can be estimated by the classical `Taylor-Culick' theory, $V_{TC}= (2 \sigma /\rho_o h)^{1/2} $, where $h$ is the film thickness. Prior to the puncture $h \sim 26 ~\mu$m, giving an estimate of the retraction velocity as $V_{TC} \sim 1.8$ m/s. The measured retraction speed of the holes in our simulations ranges from $0.75$ m/s to $1.8$ m/s. This ensures that although our simulations cannot predict the exact location of the puncture, and their formation is mesh-dependent, its expansion is well predicted, ensuring that the physics is fully-resolved following the formation of the hole. Additionally,  the  film thickness of the sheet prior to its rupture agrees with the findings of \citet{marston_jfm_2016}, \cite{Lhuissier_prl}, and \cite{Ling_prf_2017} who observed  the formation of holes in dynamics sheets of the same order of magnitude (e.g., \citet{marston_jfm_2016} and \citet{Ling_prf_2017} reported film rupture in the range of $9-16 \mu$m, and about $22\mu$m, respectively).

\subsubsection{Droplet size distribution}\label{sec:DPS}

This section draws attention to the size distribution of the droplets, formed as a result of the ruptures of the liquid threads. Attention is focused on Case 4, which is in the inertia-dominated regime characterised by high $Re$ and $We$; as discussed above, the flow in this case is accompanied by significant drop creation. During the early stages, the formation of droplets is only observed inside the leading-edge structure owing to film rupture via hole formation. At later times, both entrapped oil-droplets and injected water-droplets coexist in the computational domain. We have identified all the droplets inside  the  entire domain, and each droplet diameter is calculated through knowledge of its volume, and the assumption of a spherical shape. 

\begin{figure}    
\centering
 \includegraphics[width=1.0\linewidth]{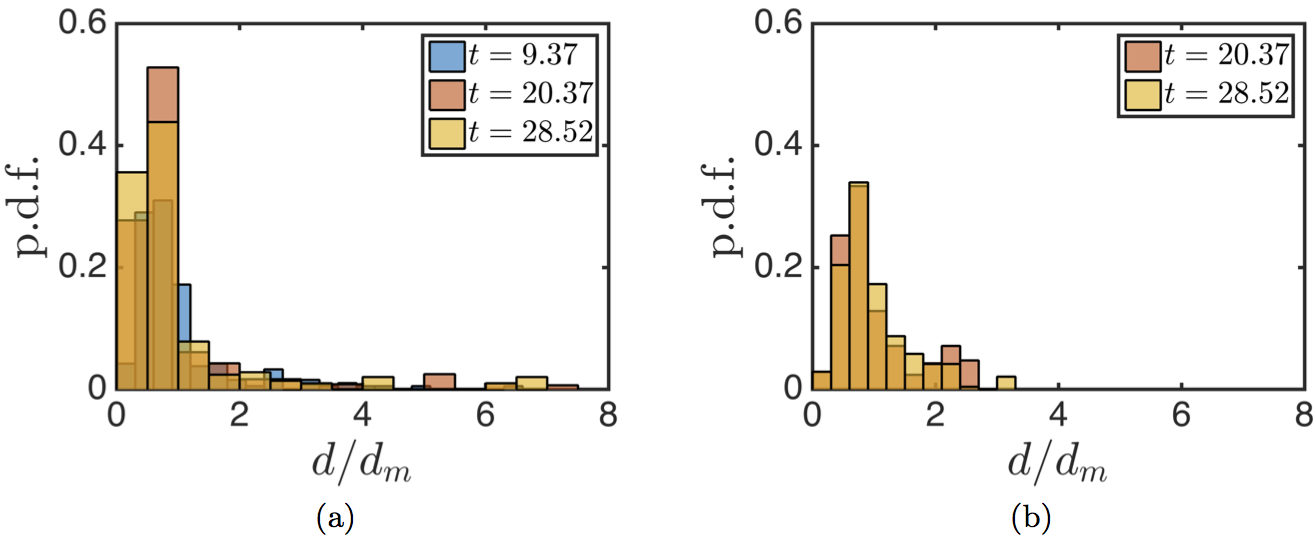}
\caption{\label{droplets} Probability density function (p.d.f.) for the entrapped oil-droplets and injected water-droplets for Case 4 ($Re=6530$ and $We=303$) at $t=9.37,~20.37$, and $28.52$ shown in panels (a) and (b), respectively. The distributions are normalised by the calculated mean diameter for each time, $d_m$.} 
\end{figure}

Figure \ref{droplets} shows the probability density function (p.d.f.) of the entrapped oil-droplets and injected water-droplets normalised by the mean droplet diameter at different temporal stages: $t=9.37,~ 20.37$ and $28.52$. The shape of the distribution for the entrapped droplets (see figure \ref{droplets}a) remains largely unchanged with increasing time except for the development of a tail; this represents the creation of larger numbers of smaller droplets due to ligament breakup as explained previously. It is possible to obtain information regarding the characteristics of the droplets in the domain; for instance, the average mean diameters of the entrapped and injected droplets are $338$ $\mu$m and $291$ $\mu$m, respectively. Figure \ref{droplets}b shows that the p.d.f. for the injected droplets has a similar shape to that of the entrapped ones depicted in \ref{droplets}a, and the distribution remains essentially unaltered for $t=20.37$ and $28.52$.

\section{Conclusions\label{conclu}}

Three-dimensional numerical simulations using a hybrid interface-tracking/level-set approach were carried out for turbulent water jets entering a stagnant and more viscous silicon oil phase. Particular attention has been paid to the temporal interfacial dynamics that arise as a result of the vortex-surface interaction. Using less computationally expensive simulations, we explored the $Re-We$ space, identifying four distinct regions based on the appearance of the interfacial structures. On this basis, we carried out high-fidelity simulations representative of each region which showed that the streamwise vorticity plays a major role in the development of the three-dimensional instabilities on the jet surface.

At low $Re$ and $We$ numbers (i.e., in the capillary-controlled regime), the jet interface remains axisymmetric and no surface corrugations are observed due to strong capillarity. Moreover, this region is also characterised by the fact that the streamwise vorticity, $\omega_x$, never becomes comparable to the azimuthal vorticity, $\omega_{\theta}$: the  azimuthal vorticity always remains two orders of magnitude larger than the streamwise vorticity. 
At low $Re$ and high $We$ numbers, the  reduction of capillarity enhances the formation of a mushroom-like structure at the leading edge of the jet, and the formation of corrugations in the jet core; however,  $\omega_{x}$ is still almost two orders of magnitude smaller than $\omega_{\theta}$. 
At high  $Re$ and low $We$ numbers, streamwise vorticity gains in importance resulting in further deformation of the jet core. 

In the inertia-controlled regime (i.e., for high $Re$ and $We$ numbers), the streamwise vorticity becomes comparable to the azimuthal vorticity  triggering the streamwise deformation of KH vortices. We show the formation of hairpin vortices near the interface, which are aligned in the streamwise direction forming layers of inner- and outer- hairpin-vortices. The formation of inner and outer lobes in the jet core are closely linked to their neighbouring hairpin-vortices. The alignment of vortices enhances the perforation of the lobes to form holes which expand radially by capillarity, and ultimately give rise to the formation of droplets. Another main feature of the inertia-controlled regime is the creation of a thin toroidal sheet inside of the leading-edge structure. The thinning of this sheet leads to its perforation to form droplets. The droplets rotate as a result of their interaction with the vorticity field,  and further topological transitions occur (e.g., coalescence). 

While in the present study we have focused on the spatial and temporal development of the interfacial dynamics as a result of the  vortex-surface interaction, further investigations should be carried out on the statistical response of the turbulent multiphase flow, similar to the work presented by \citet{Ling_jfm_2019} for a mixing layer between parallel gas and liquid streams. Additionally, our study assumes a constant value of surface tension, but it is known that streams are usually contaminated with deliberately-placed or naturally-occurring surfactants, which reduce the surface tension and give rise to surface tension gradients and Marangoni stresses. As shown recently by \citet{constanteamores2020dynamics}, surfactants are capable of inhibiting capillary singularities and rigidifying the interface, subsequently changing the fate of the atomisation. Hence, the presence of surfactants may change the present results, providing us with an exciting avenue of future research.  \\

Declaration of Interests. The authors report no conflict of interest. \\

\subsection*{Acknowledgements}

This work is supported by the Engineering $\&$ Physical Sciences Research Council, United Kingdom, through a studentship for C.R C-A. in the Centre for Doctoral Training on Theory and Simulation of Materials at Imperial College London funded by the EPSRC (Grant No. EP/L015579/1), and through the EPSRC MEMPHIS (Grant No. EP/K003976/1) and PREMIERE (EP/T000414/1) Programme Grants. The authors would like to acknowledge the funding and technical support from BP through the BP International Centre for Advanced Materials (BP-ICAM), which made this research possible. O.K.M. also acknowledges funding from PETRONAS and the Royal Academy of Engineering for a Research Chair in Multiphase Fluid Dynamics, and the PETRONAS Centre for Engineering of Multiphase Systems. D.J. and J.C. acknowledge support through computing time at the Institut du Developpement et des Ressources en Informatique Scientifique (IDRIS) of the Centre National de la Recherche Scientifique (CNRS), coordinated by GENCI (Grand Equipement National de Calcul Intensif) Grant No. 2020A0082B06721.  We also acknowledge HPC facilities provided by the Research Computing Service (RCS) of Imperial College London for the computing time. The numerical simulations were performed with code BLUE \citep{Shin_jmst_2017} and the visualisations have been generated using ParaView.

\section*{Appendices }

\subsection*{Mesh study and resolution considerations}

Solving the small scales of the atomisation phenomenon 
is a challenging process, and in order to ensure the physical validity of the results, we have assessed the grid dependence nature of our results by performing a mesh study for Case 4 (see table \ref{tab:mesh}). The appropriate choice for the minimum grid length-scale will depend on either  the vanishing interfacial singularity or the smallest turbulence length-scale (i.e., Kolmogorov-length-scale).

For single-phase jets,  \cite{Pope_2000} suggested that the Kolmogorov-length-scale  $\eta$ is resolved if $\Delta x / \eta  \leq 2.1$,  where $\Delta x$ represents the minimum computational cell size, and $\eta$ is estimated by $\eta = Re^{-3/4} l$, where $l$ is the length scale. Previous computational studies of temporal  and spatial multiphase jets  are presented in table \ref{re_studies}, showing the current state-of-the-art in terms of Pope's criterion. It is evident that Pope's criterion is difficult to meet due to  the high computational costs of the simulations.  In their recent work, \cite{Ling_prf_2017,Ling_jfm_2019} performed simulations of a two-phase mixing layer  between parallel gas  and liquid streams to investigate the interfacial dynamics and the statistics of the multiphase turbulence, estimating the Kolmogorov length scale to be $\eta \sim 0.945  \mu m$, which leads to  $\Delta x / \eta \sim 3$. However, through their numerical simulations, they estimated that $\eta$ is larger in size  (e.g., $\eta= 3- 4.5 \mu$m), and they also showed that  their lower resolution mesh, i.e., $\Delta x / \eta \sim 6$ (see their figure 19d), was capable of predicting similar results in terms of turbulence dissipation.

For our highest Reynolds number, our simulation does not meet Pope's criterion. But as shown by \citet{Ling_jfm_2019}, the actual $\eta$ could be larger.  Additionally, the atomisation of the injected phase in a stagnant viscous  phase alleviates the range of relevant physical scales.

\begin{table}
\begin{center}
\caption{\label{re_studies}List of computational studies of atomisation showing compliance  with the Pope criterion \citep{Pope_2000}. }
\begin{tabular}{ c c c c c c }
 Study & Configuration & Type of study &   Re& $\Delta x / l$  &$\Delta x / \eta$ \\ 
 \hline
 \citet{Desjardins_as_2010}  &Planar& Temporal & $2000-3000$ & $0.015$  & $\sim 5-6$    \\ 
 \citet{Hermann_as_2011} &Round & Spatial & $5000$ & $0.031$/$0.0078$  & $\sim 5-19$    \\ 
   \citet{Jarrahbashi_jfm_2016}  &Round& Temporal & $1600-16000$ & $0.0125$  & $\sim 3-18$    \\  
   \citet{Zandian_2016,Zandian_jfm_2018}  &Planar& Temporal & $2500-5000$ & $0.025$  & $\sim 9-15$    \\  
   \citet{zandian_sirignano_hussain_2019}  &Round& Spatial & $2000-3200$ & $0.01$  & $\sim 3-4$    \\  
   \citet{Ling_jfm_2019}  &Planar& Temporal & $8000 $& $0.0039$  & $\sim 3$    \\  
   Current study  &Round& Spatial & $1000-6530$ & $0.0065$  & $\sim 1-4.5$    \\  
\end{tabular}
\end{center}
\end{table}

\begin{table}
\centering
\caption{\label{tab:mesh} Characteristics of different mesh sizes used to study the jet dynamics in this study.}
\begin{tabular}{ccccc}
~ Run ~  & \begin{tabular}[c]{@{}l@{}}~ Global mesh size\\~ (number of cells) ~ ~\end{tabular} & \begin{tabular}[c]{@{}l@{}}~ Number of parallel ~\\~ ~process threads ~ ~\end{tabular} & \begin{tabular}[c]{@{}l@{}}~ Minimum mesh\\~~~~~~ size ($\mu$m) ~ ~\end{tabular} & \begin{tabular}[c]{@{}l@{}}~~~ Total~Comput.\\ ~~~~hours per CPU\end{tabular}  \\
\hline
 M1		&  		 $768\times192\times192$                         & $12 \times 3 \times 6 =72 $                &$104.1$              & $ \sim 180$          \\
 M2		&	         $1536\times192\times192 $                       & $24 \times 6 \times 6 =864$                & $58.5 $              & $\sim 100$           \\
 M3		&		$3072\times384\times384$                        & $ 48 \times  6 \times  6 = 1728$                 & $26.1$               & $\sim 500$                                                                   
\end{tabular}
\end{table}

The second biggest challenge of computational atomisation is `numerical breakup' of liquid threads. A coarse grid would trigger the formation of thicker numerical threads, and consequently the formation of larger droplets (this problem has been previously reported by  \citet{Shinjo_2010}, \citet{Hermann_as_2011}, \citet{Gorokhovski_arfm_2008}, and \citet{Jarrahbashi_pof_2014}). \citet{Shinjo_2010} stated that the mesh should be refined up to the point where the dynamics of the thread are solely governed by surface tension forces. These forces would trigger the formation of capillary waves during the retraction of these ligaments, giving rise to the onset of the Rayleigh-Plateau instability (i.e., the `end-pinching' mechanism). After those mechanisms are initialised, the thread-dynamics enter in an asymptotic behaviour towards the interfacial singularity (i.e., a refined mesh would not affect the size of the resulting droplets). On this basis, the numerical resolution regarding the interfacial length scales will be assessed following the methodology proposed  by \citet{Menard_ijmf_2007} and \citet{Desjardins_as_2010}, who used a `grid-based Weber number', defined by $We_{\Delta x_{min}} = \rho_{_w} {U^2}\Delta x_{min} / \sigma $. This equation provides us with the smallest interfacial length scale that the simulation is capable of resolving by assuming that the smallest interfacial structure is equal to the minimum mesh size of the computational domain. \cite{Menard_ijmf_2007} suggested that no further breakup is observed for values under $10$. For a Reynolds number corresponding to $Re \sim 10^4$ and M3-type mesh, the grid-based Weber number is $We_{\Delta x_{min}} \sim 4.12$, which meets the above criterion, suggesting that all capillary singularities would be resolved.

\begin{figure}
\centering
 \includegraphics[width=1.0\linewidth]{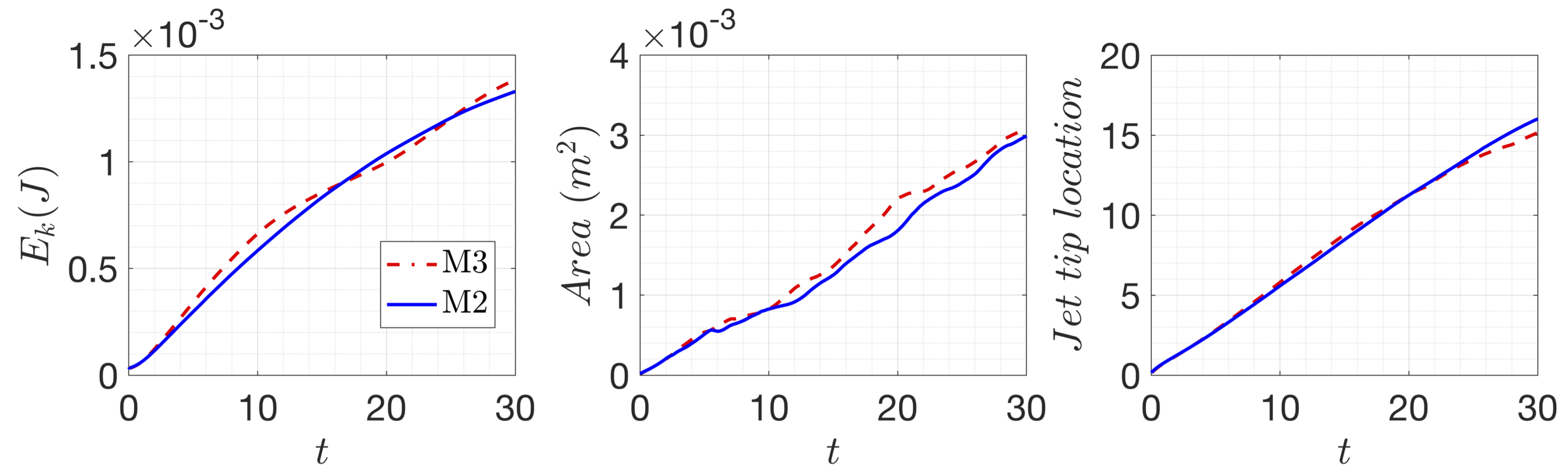}
\caption{ \label{fig:mesh} Mesh study for Case 4 ($Re=6530$ and $We=303$). The panels highlight the temporal evolution of the kinetic energy $E_k$ (left panel),  the interfacial area (middle panel) and the maximum axial location of the jet tip (right panel).}
\end{figure}

Therefore, we have proved that the M3 mesh is capable of resolving turbulent-scales and interfacial singularities, and consequently detailed analysis of interfacial and vortical structures is performed using a M3 mesh-type (unless stated otherwise).  Figure \ref{fig:mesh} shows the temporal evolution of the kinetic energy,  $E_k=\int_V (\rho \textbf{u}^2)/2 ~\rm{d} V$, the interfacial area and the maximum axial location of the jet tip (i.e., leading edge)  for different meshes. Additionally, the vorticity profiles were checked between the M2- and M3-mesh, and no significant differences were found.

\subsection*{Linear stability analysis}

Following  \cite{Plateau_1873} and \cite{Rayleigh_1879}, we will show that our numerical method is capable of predicting the  growth rates in relation to the capillary breakup of jets (i.e., Rayleigh-Plateau instability).  On this basis, we have considered an inviscid  cylindrical jet of radius $R_o$ and  density $\rho$,  surrounded by a dynamically passive gas.  We consider the evolution of infinitesimal perturbations to linearise the equations. Thus, the perturbed jet is represented by 

\begin{equation}
R(\omega,\varepsilon)=R_o \left (1+\varepsilon \cos\left ( \omega x\right )\right )= R_o \left (1+\varepsilon \cos\left ( \frac{2\pi }{\lambda } x\right )\right )
\end{equation}

\noindent
where $\varepsilon$, $\lambda$, $\omega$ stand for the amplitude of infinitesimal perturbation, wavelength and the growth rate, respectively. After the linearisation of the governing equations  and retaining terms only to order of $\varepsilon$ results in a dispersion relation, which indicates the dependence of the growth rate on the wavenumber (i.e., $k=2 \pi/\lambda$) of the inviscid jet, expressed as :

\begin{equation}
\label{RP_dispersion}
\omega^2=\frac{\sigma \left(kR_o\right) }{\rho R_o^3}\frac{I_1(kR_0)}{I_0(kR_0)}\left(1-k^2R_o^2\right)=\omega_o^2 x \frac{I_1(x)}{I_0(x)} \left(1-x^2\right)
\end{equation}

\noindent
where $x=kR_o$ stands for a reduced wavenumber and $\omega_o^2=\sigma/(\rho R_o^3)$. When $x>1$, the initial perturbation presents stable oscillatory solutions; whereas in the other case (i.e., $x<1$), the initial perturbation grows exponentially with a growth rate given by equation (\ref{RP_dispersion}). 

\begin{figure}  
\centering
 \includegraphics[width=0.6\linewidth]{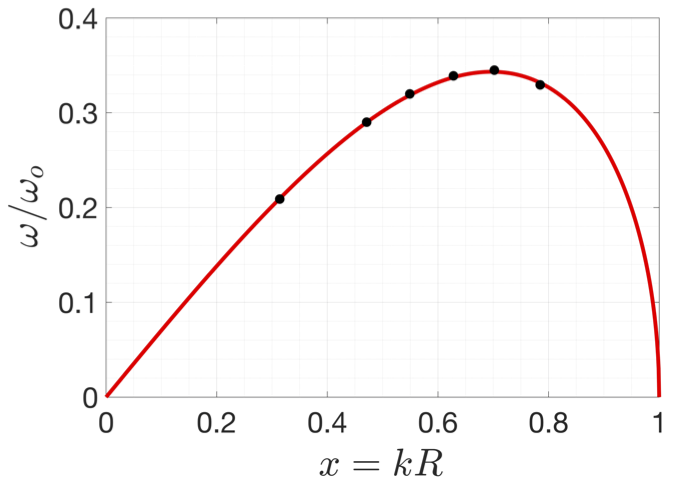}
\caption{\label{w_rates} Comparison of nondimensional growth rates between numerical simulations (black markers)  and linear theory (solid line).}
 \end{figure}

Figure \ref{w_rates} shows the  growth rates  obtained through the numerical simulations in caparison to the  the theoretical dispersion relation. The growth rates from direct numerical simulations (DNS) shows a discrepancy under $2\%$ in all cases in comparison to the theoretical values. Based on this, we can conclude the capability of our numerical method for capturing accurately  the dynamics of capillary-jets.

\subsection*{Scalings laws for the capillary breakup of liquid threads}\label{sec:RP_section}

As we have presented in the introduction, direct numerical simulations must be capable of predicting the developing two-phase fluid interfacial dynamics featuring interface breakup, and droplet coalescence. In light of this, we aim to show the capabilities of our numerical framework in the prediction of the scaling laws for the capillary singularity of liquid threads.  As the point of singularity approaches, the system is driven locally by the large interfacial curvature, and its interfacial dynamics depends solely upon the physical properties of the liquid.  \citet{Lister_Stone_pf_1998} have suggested that the pinchoff of a viscous thread (of radius $r(z,t)$, density $\rho$, viscosity $\mu$ and surface tension $\sigma$) surrounded by another viscous fluid  transits between different dynamical regimes as $r(t) \rightarrow  0$ towards the breakup time $\tau$ (see figure 1 of \citet{Lister_Stone_pf_1998}). The thinning of a water thread, surrounded by air, transitions from an inertial-capillary regime ($r\sim \tau^{2/3} $ and $u\sim \tau^{-1/3}$ ) to an inertial-viscous regime ($r\sim \tau $ and $u\sim \tau^{-1/2}$ ), more details can be found in \citet{Day_prl_1998}, \citet{Eggers_prl_1993} and \citet{Lister_Stone_pf_1998}.

\begin{table}
\centering
\caption{Characteristics of different mesh sizes used to study the capillary breakup of a water thread.}
\begin{tabular}{ccccc}
~ Run ~  & \begin{tabular}[c]{@{}l@{}}~ Global mesh size\\~ (number of cells) ~ ~\end{tabular} & \begin{tabular}[c]{@{}l@{}}~ Number of parallel ~\\~~ ~process threads ~ ~\end{tabular} & \begin{tabular}[c]{@{}l@{}}~ Pinch-off time\\~~~~~~~~~(s) ~ ~\end{tabular} & \begin{tabular}[c]{@{}l@{}}~~~ Total~Comput.\\ ~~~~hours per CPU\end{tabular}  \\
\hline
~ V1 ~ & $64\times64\times256$        & $16$   & $0.375$       & $\sim 2$        \\
~ V2 ~    & $96\times96\times384$          & $54$   & $0.372$      & $\sim 20 $     \\
~ V3 ~    & $192\times192\times768$        & $432$ &  $0.369$     & $ \sim 48$                                                                   
\end{tabular}
\label{tab:mesh_pinchoff}
\end{table}

\begin{figure}
\centering
 \includegraphics[width=1.0\linewidth]{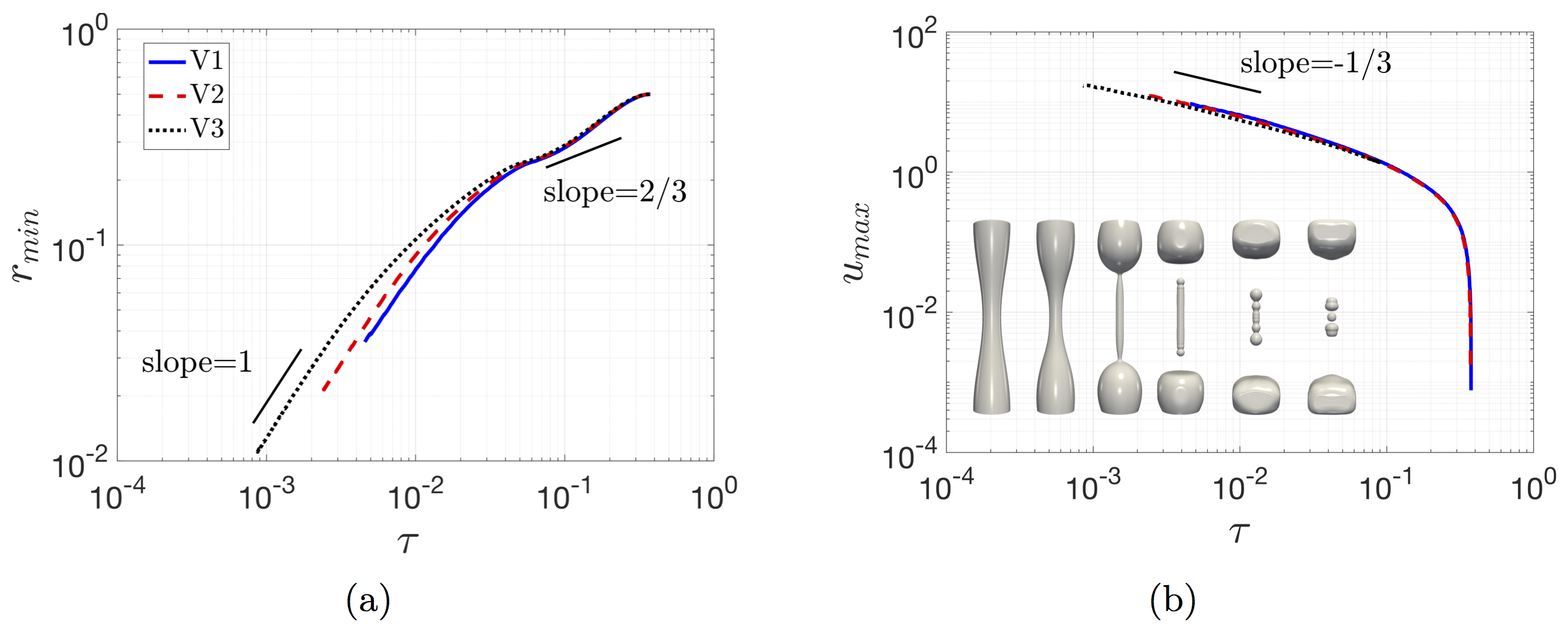}
\caption{ \label{fig:pinchoff} Scaling laws for the capillary singularity of a water-thread surrounded by air for different numerical resolutions (see table \ref{tab:mesh_pinchoff}). Here, the Ohnesorge number is $1.178 \times 10^{-3}$. The minimum thread radius, (a), and the maximum streamwise velocity, (b), versus the remaining time to breakup $\tau$, agree with the inertial-capillary and inertial-viscous regimes presented by \citet{Eggers_prl_1993,Day_prl_1998}. 
Additionally,  a three-dimensional representation of the interface is shown to depict the retraction of the satellite ligament to produce daughter droplets via the `end-pinching' mechanism (in agreement with \citet{notz_basaran_2004}).}
\end{figure}

Therefore, we consider the capillary singularity of a water thread (of initial radius $R$) surrounded by air in absence of gravity as the interfacial dynamics are locally driven by a balance of viscous-capillary and inertial forces, which can be expressed by the  Ohnesorge number $Oh=\mu/(\rho \sigma R)$. As $Oh <1$, at early thinning stages there is a competition between the fluid inertia and the opposing capillary pressure, and the predicted thread radius towards the singularity agrees with the theoretical scalings (see figure \ref{fig:pinchoff}) as the thread transits between  different regimes (in agreement with  \citet{Castrejon_pnas_2015}). After the singularity point, the formation of a satellite ligament is observed which has an initially cylindrical shape (with $Oh=2.62 \times 10^{-3} $ and $L_o=11.55$); however,  it undergoes retraction driven by capillarity to form more spherical droplets. This process known as `end-pinching' has been previously well described by \citet{notz_basaran_2004}. By performing this analysis we have  proved the capability of our numerical technique  to predict the dynamics of the capillary singularity of a liquid thread and its post-breakup events (table \ref{tab:mesh_pinchoff} shows a summary of the different meshes evaluated for this study).

Additionally, the accuracy and validation of the numerical method has been previously addressed  to other complex interfacial phenomena. These phenomena include  breakup and recoiling of liquid threads \citep{constanteamores2020dynamics}, falling film flows \citep{batchvarov2020threedimensional}, propagation of elongated bubbles in channels \citep{batchvarov2020effect}, bubbles undergoing bursting \citep{constanteamores2020bb}, and drops coalescing partially or completely with deformable interfaces.

\subsection*{Time step}
Finally, the temporal integration scheme is based on a second-order Gear method, with implicit solution of the viscous terms of the velocity components. The time step $\Delta t$ is adaptive to ensure stability, and it is defined through the criterion:

\begin{equation}
\Delta t = \min \left\{\Delta t_{cap},~\Delta t_{vis},~\Delta t_{CFL}, ~\Delta t_{int}\right\} 
\end{equation}

\noindent
where  $\Delta t_{cap}$, $\Delta t_{vis} $, $\Delta t_{CFL}$, $ \Delta t_{int}$ stand for the capillary time step, the viscous time step, the Courant- Friedrichs-Lewy (CFL) time step, and interfacial CFL time step, respectively. Those terms are defined by

\begin{equation} 
\begin{matrix}
\Delta t_{vis} = \min \left (  \dfrac{\rho_{_w}}{\mu_{_{w}}}, \dfrac{\rho_{_{so}}}{\mu_{_{so}}}\right ) \dfrac{ \Delta x^2_{\min} }{6  }
 & ,~
\Delta t_{cap} =\dfrac{1}{2} \left (  \dfrac{(\rho_{_w} +\rho_{_{so}}) \Delta {x_{\min}}^3}{\pi \sigma }\right )^{1/2} ,\\
& ,~
\\
\Delta t_{CFL} = \underset{j}{\min} \left (  \underset{domain}{\min} \left (   \dfrac{ \Delta x_{j} }{u_j} \right )\right ) 
 & ,~
\Delta t_{int} = \underset{j}{\min} \left (  \underset{\Gamma (t)}{\min}  \left ( \ \dfrac{ \Delta x_{j} }{\left \| \mathbf{V} \right \|}\right )\right ),
\\
&
\end{matrix}
\end{equation}

\noindent
where $\Delta {x_{\min}} = \min_j (\Delta x_j)$, respectively. In our simulations the adaptive time-step is controlled  by $\Delta t_{int}$ which is $O(10^{-5})$s.


\end{document}